\documentclass[a4paper,11p]{article}

\pdfoutput=1

\usepackage{jheppub} % for details on the use of the package, please
\usepackage[T1]{fontenc} 
\usepackage{graphicx}
\usepackage{epsfig}
\usepackage{amssymb}
\usepackage{amsfonts}
\usepackage{amsmath,euscript,array,mathrsfs}
\usepackage{bbold}
\usepackage{epsf}
\usepackage{slashed}
\usepackage{comment}
\usepackage{colortbl}

\usepackage[caption=false]{subfig}
\usepackage[colorlinks=true,linktocpage=true,linkcolor=blue,citecolor=blue]{hyperref}

\usepackage{tikz}
\usetikzlibrary{decorations.markings,decorations.pathmorphing}

\newcommand{\be}{\begin{equation}}
\newcommand{\ee}{\end{equation}}
\newcommand{\beqs}{\begin{eqnarray}}
\newcommand{\eeqs}{\end{eqnarray}}

\newcommand{\U}{\mathcal{U}}

\makeatletter
\gdef\@fpheader{}
\makeatother
\begin{document}

\title{Evolutions in first-order viscous hydrodynamics}

%\author{Hans Bantilan$^1$}
%\author{Yago Bea$^{1,2}$}
%\author{Pau Figueras$^1$}

%\vspace{1mm}

%\affiliation{$^1$ School of Mathematical Sciences, Queen Mary University of London, Mile End Road, London E1 4NS, United Kingdom}

%\vspace{1mm}

%\affiliation{$^2$ Department of Physics and Helsinki Institute of Physics, PL 64, FI-00014 University of Helsinki, Finland}

\author[a]{Hans Bantilan}
\author[b]{Yago Bea}
\author[a]{Pau Figueras}

\affiliation[a]{School of Mathematical Sciences, Queen Mary University of London, Mile End Road, London E1 4NS, United Kingdom}
\affiliation[b]{Department of Physics and Helsinki Institute of Physics, PL 64, FI-00014 University of Helsinki, Finland}

% e-mail addresses: one for each author, in the same order as the authors
\emailAdd{h.bantilan@qmul.ac.uk}
\emailAdd{yago.beabesada@helsinki.fi}
\emailAdd{p.figueras@qmul.ac.uk}

\abstract{
Motivated by the physics of the quark-gluon plasma created in heavy-ion collision experiments, we use holography to study the regime of applicability of various theories of relativistic viscous hydrodynamics. Using the microscopic description provided by holography of a system that relaxes to equilibrium, we obtain initial data with which we perform real-time evolutions in 2+1 dimensional conformal fluids using the first-order viscous relativistic hydrodynamics theory of Bemfica, Disconzi, Noronha and Kovtun (BDNK),  BRSSS and ideal hydrodynamics.  By initializing the hydrodynamics codes at different times, we can check the constitutive relations and assess the predictive power and accuracy of each of these  theories.
}

\maketitle

%%%%%%%%%%%%%%%%%%%%%%%%%%%%%%%%%%%%%%%%%%%%%%%%
\section{Introduction}
%\label{intro}
%%%%%%%%%%%%%%%%%%%%%%%%%%%%%%%%%%%%%%%%%%%%%%%%
Relativistic hydrodynamics is a very general theory that provides the effective description of the out-of-equilibrium real-time dynamics of many physical systems of interest. 
In fact, hydrodynamics is better understood as a low energy effective field theory (EFT) constructed as a derivative expansion near a local thermodynamic equilibrium. As such, the fundamental variables are the basic quantities describing the equilibrium solution, such as energy density and velocities, and their gradients. The zeroth order piece in the gradient expansion corresponds to ideal hydrodynamics. 

The equations of ideal hydrodynamics are hyperbolic, having a well-posed initial value problem for general initial conditions, and are amenable to numerical simulations.
However,  there are important physical situations where dissipative effects may play a relevant role, such as in the Quark-Gluon Plasma (QGP)  \cite{Romatschke:2017ejr,Busza:2018rrf} or in certain astrophysical systems \cite{Alford:2017rxf,Chabanov:2021dee,Shibata:2017jyf,Fujibayashi:2017puw,Fujibayashi:2020jfr}.
However, it is well-known that if first-order viscous terms are included in the relativistic hydrodynamic expansion (in the Landau frame), the resulting equations of motion  are parabolic and hence incompatible with the causality postulates of relativity \cite{Hiscock:1985zz}.

A well-known approach to address this issue of lack of hyperbolicity and causality is exemplified by the work of M\"uller, Israel and Stewart (MIS) \cite{Muller:1967zza,Israel:1979wp,Israel:1976tn}. The strategy consists in including second order terms and introduce new variables with the respective equations of motion, such that the overall system of equations is hyperbolic and causal. The new theory should have the same IR, i.e.,  hydrodynamic regime, but different UV properties.\footnote{It is not clear that such modification of the theory correctly describes the physics in situations where there is a strong UV cascade, such as in a 3+1 turbulent flow.}
The MIS formulation is not unique; there are several formulations which include extra variables and equations to recover hyperbolicity: BRSSS \cite{Baier:2007ix}, DNMR \cite{Denicol:2012cn}, divergence type theories \cite{Geroch:1990bw,Lehner:2017yes}, etc.. As discussed in \cite{Geroch:1995bx,Geroch:2001xs}, in principle  all these theories should be equivalent in the IR.

%BDNK
In recent years it has been realized that it is possible to write the equations of first-order viscous hydrodynamics in a way such that they are hyperbolic, without the need of adding extra variables or equations. When including first order terms in the hydrodynamic expansion, the basic hydrodynamic variables present an ambiguity when defined out of equilibrium, and fixing this ambiguity is known as choosing a `frame'. Traditionally, it was common to use only natural frames such as the Landau frame. The insight of BDNK \cite{Bemfica:2017wps,Bemfica:2019knx,Kovtun:2019hdm} was to notice that the choice of frame affects the hyperbolicity of the equations of motion and a certain frame can be chosen such that the latter are hyperbolic.\footnote{See \cite{Armas:2022wvb} for a recent generalization to magnetohydrodynamics.}

MIS-based hydrodynamic codes have been extensively used for many years to study dissipative effects. Given that there is now a new theory of viscous relativistic hydrodynamics, it is natural to explore it. 
The first ones to do so were the authors of \cite{Pandya:2021ief}, who studied the evolution of smooth and non-smooth initial data using BDNK in effectively 1+1 dimensions.  The purpose of the present paper is to carry out the first studies of BDNK focusing on applications to the QGP. More concretely, in this article we report on studies of dynamical evolutions of the BDNK equations in a 2+1 dimensional conformal uncharged fluid. 
In order to acquire a better understanding of BDNK as a theory of viscous hydrodynamics, we quantitatively compare it to ideal and  BRSSS \cite{Baier:2007ix} hydrodynamics, as well as to the microscopic theory provided by holography.

%Holography.
The QGP created in heavy-ion collision experiments is initially far from equilibrium and it subsequently relaxes to a hydrodynamic regime. 
 Hydrodynamic numerical codes are used to provide effective descriptions of these evolutions, where the underlying microscopic theory is Quantum Chromodynamics (QCD).  
Motivated by this picture, in this paper we use numerical general relativity and holography to obtain a first principles description of the microscopic real-time dynamics of the stress tensor in a strongly coupled field theory of a system which relaxes to equilibrium. 
The microscopic solution is valid for all times, covering both the far-from-equilibrium  and near equilibrium hydrodynamic regimes. Whilst previous studies of the applicability of hydrodynamics in holographic systems have focused on checking the constitutive relations, in this paper we go beyond the state of the art and use the microscopic data to initialize the hydrodynamic codes at different times \footnote{Previous studies in a one-dimensional setting were performed in \cite{Heller:2014wfa}.}. In this way, by comparing with the UV-complete solution provided by holography, we verify both the constitutive relations and the predictivity of each hydrodynamic theory for a certain class of initial conditions.  We use units $c=G_4=1$.

When our article was nearing completion,  \cite{Pandya:2022pif} appeared on arXiv, extending the numerical studies of the BDNK equations.

%%%%%%%%%%%%%%%%%%%%%%%%%%%%%%%%%%%%%%%%%%%%%%%%

\section{Hydrodynamics: the equations} 
%\label{initial}
%%%%%%%%%%%%%%%%%%%%%%%%%%%%%%%%%%%%%%%%%%%%%%%%
We use three sets of hydrodynamic evolution equations: ideal hydrodynamics, BRSSS and BDNK. We write these equations for a $2+1$ dimensional conformal fluid in Minkowski spacetime. Conformal symmetry fixes the equation of state: 
\begin{equation}
p = \frac{\epsilon}{2}\,,  
\label{eos_conformal}
\end{equation}
where $\epsilon$ is energy density and  $p$ is pressure. The temperature $T$ is related to the energy density as $\epsilon=\frac{2}{3} A\,T^{3}$, where  $A$ is an arbitrary constant that in our simulations we fix to $A=4\pi^2/9$.\footnote{For a holographic fluid in $d$ spacetime dimensions, this constant $A$ is related to the bulk's Newton's constant of gravitation $G_{d+1}$ as $A=\frac{(4 \pi )^d}{16 \pi \,G_{d+1} d^{d-1}}$.} Considering the 3-velocity of the fluid $u^\mu$, normalized such that $u^2=-1$,  we define $\dot{\epsilon}\equiv u^{\mu}\nabla_{\mu}\epsilon$, $\nabla_{\perp}^{\mu}\equiv \Delta^{\mu\nu} \nabla_{\nu} $, $\nabla \cdot  u \equiv \nabla_{\rho} u^{\rho}$ , $\Delta^{\mu\nu} \equiv g^{\mu\nu} + u^\mu\,u^\nu$, $\Omega^{\mu\nu} \equiv \left( \nabla^{\mu}_{\perp} u^{\nu} - \nabla^{\nu}_{\perp} u^{\mu} \right)/2$, $\sigma^{\mu\nu} \equiv 2  \nabla^{\langle\mu} u^{\nu\rangle}$, where
\begin{align}
&A^{\langle\mu\nu\rangle} \equiv \frac{1}{2} \Delta^{\mu\alpha} \Delta^{\nu\beta}(A_{\alpha \beta}+A_{\beta \alpha})-\frac{1}{2}\Delta^{\mu\nu} \Delta^{\alpha\beta}A_{\alpha \beta}  \,, %\\
\end{align}
is symmetric, traceless and transverse to $u_{\mu}$. The constitutive relations of relativistic hydrodynamics up to second order in the derivative expansion in the Landau frame can be written as \cite{Baier:2007ix},
\begin{subequations}
\begin{equation}
T^{\mu\nu} = \epsilon\,u^\mu\,u^\nu + p\,\Delta^{\mu\nu} + \Pi^{\mu\nu}\,,
\label{constitutive0} 
\end{equation}
with
\begin{align}
& \Pi^{\mu\nu}=-\eta\, \sigma^{\mu\nu}+\eta\,\tau_{\pi} \left(\dot{\sigma}^{\langle\mu\nu\rangle}+\frac{1}{2} \,\sigma^{\mu\nu} \,\nabla \cdot  u\right) \nonumber \\
& +{\lambda_1}\, {\sigma^{\langle\mu}}_{\rho}\sigma^{\nu\rangle\rho}+{\lambda_2}\, {\sigma^{\langle\mu}}_{\rho}\Omega^{\nu\rangle\rho}+\lambda_3\, {\Omega^{\langle\mu}}_{\rho}\Omega^{\nu\rangle\rho} \,,
\label{sheartensor0} 
\end{align}
\label{constitutive0sheartensor0} 
\end{subequations}
where $\eta$ is the shear viscosity and $\tau_{\pi}$, $\lambda_1$, $\lambda_2$, $\lambda_3$ are second order transport coefficients. The constitutive relations of ideal hydrodynamics are given by \eqref{constitutive0} with $\Pi^{\mu\nu}=0$. 

The conservation of the stress-energy tensor
\begin{equation}
\nabla_{\mu}T^{\mu\nu} = 0\,, 
\label{conservation0} 
\end{equation}
provides the evolution equations for the dynamical variables $\epsilon$ and the independent components of the velocity vector $u^\mu$. The equations of ideal hydrodynamics are obtained by plugging (\ref{constitutive0}) with $\Pi^{\mu\nu}=0$  into (\ref{conservation0}).

The equations of  BRSSS \cite{Baier:2007ix} are obtained from \eqref{constitutive0sheartensor0} and \eqref{conservation0} by promoting $\Pi^{\mu\nu}$ to a new variable; then, using the first order relation $\Pi^{\mu\nu}=- \eta \sigma^{\mu\nu} $, equation \eqref{sheartensor0} becomes an independent evolution equation for $\Pi^{\mu\nu}$:
\begin{align}
&\Pi^{\mu\nu}=-\eta \sigma^{\mu\nu}-\tau_{\pi} \left(\dot{\Pi}^{\langle\mu\nu\rangle}+\frac{3}{2} \Pi^{\mu\nu} \nabla \cdot  u\right) \nonumber \\
&+\frac{\lambda_1}{\eta^2} {\Pi^{\langle\mu}}_{\rho}\Pi^{\nu\rangle\rho}-\frac{\lambda_2}{\eta} {\Pi^{\langle\mu}}_{\rho}\Omega^{\nu\rangle\rho}+\lambda_3 {\Omega^{\langle\mu}}_{\rho}\Omega^{\nu\rangle\rho} \,.
\label{BRSSS1}
\end{align}

We now present the BDNK equations. Given a timelike unit vector $u^{\mu}$, a symmetric tensor can be decomposed as
\begin{equation}
T^{\mu\nu} = \mathcal{E}\,u^\mu\,u^\nu + \mathcal{P}\,\Delta^{\mu\nu} + (\mathcal{Q}^\mu\,u^\nu + \mathcal{Q}^\nu\,u^\mu) + \mathcal{T}^{\mu\nu}\,, 
\end{equation}
where
 \begin{subequations}
	\begin{align}
	\mathcal{E} & \equiv u_\mu\,u_\nu\, T^{\mu\nu}  \,, \\
	\mathcal{P} & \equiv \frac{1}{2} \Delta_{\mu\nu} T^{\mu\nu}  \,,\\
	\mathcal{Q}_{\mu} & \equiv -\Delta_{\mu\alpha} u_{\beta} T^{\alpha\beta}  \,,\\
	\mathcal{T}_{\mu\nu} & \equiv \frac{1}{2} \left( \Delta_{\mu\alpha} \Delta_{\nu\beta}+ \Delta_{\nu\alpha}\Delta_{\mu\beta}-\Delta_{\mu\nu}\Delta_{\alpha \beta} \right)  T^{\alpha\beta}  \,.
	\end{align}
\end{subequations}
For a conformal fluid, the expansion in derivatives of each component to first order is 
 \begin{subequations}
 	\begin{align}
 	\mathcal{E} &= \epsilon + 2 \pi_2 \left(\frac{2}{3} \frac{\dot \epsilon}{\epsilon} + \,\nabla \cdot u \right) \,, \\
 	\mathcal{P}  &= p + \pi_2\, \left(\frac{2}{3} \frac{\dot \epsilon}{\epsilon} + \,\nabla \cdot u \right) \,,\\
 	\mathcal{Q}^{\mu} &= \theta_1 \left( \dot{u}^{\mu}+\frac{1}{3} \frac{\nabla_{\perp}^{\mu} \epsilon}{\epsilon}  \right) \,,\\
 	\mathcal{T}^{\mu\nu} & = -\eta\,\sigma^{\mu\nu} \,,
 	\end{align}
 \end{subequations}
where $\pi_2$, $\theta_1$ and $\eta$ are transport coefficients.
 Conformal symmetry fixes the dimensions of $\pi_2$ and $\theta_1$, and we write them as a constant times the shear viscosity $\eta$
 \begin{equation}
 \pi_2 = a_2 \,\eta \,, ~~~~  \theta_1 = a_1\, \eta   \,. 
 \end{equation}
 The stress tensor, to first order in the derivative expansion, is then
\begin{equation}
\begin{aligned}
T^{\mu\nu} =& \left[\epsilon + 2\, a_2 \eta\left( \frac{2}{3}\frac{\dot{\epsilon}}{\epsilon}+\nabla \cdot u \right)\right]\left(u^\mu\,u^\nu + \frac{\Delta^{\mu\nu}}{2}\right) \\
&+a_1 \eta \left[\left(\dot{u}^{\mu}+\frac{1}{3} \frac{\nabla_{\perp}^{\mu} \epsilon}{\epsilon} \right)u^\nu + (\mu\leftrightarrow \nu)\right]\\
&-\eta\,\sigma^{\mu\nu}  + O(\partial^2)\,.
\end{aligned}
\label{eq:tmunu11}
\end{equation}
The BDNK equations are obtained from plugging \eqref{eq:tmunu11} into the conservation equation \eqref{conservation0}. 

The constants $a_1$ and $a_2$ specify the frame. The Landau frame corresponds to $a_1=a_2=0$, by which we recover (\ref{constitutive0sheartensor0}) up to first order terms. The BDNK equations are hyperbolic iff  \cite{Bemfica:2019hok}
\begin{equation}
a_2>1  \,, ~~~~   a_1>\frac{4\,a_2}{a_2-1}  \,. 
\label{gap}
\end{equation}
The Landau frame lies outside this region, and there is a gap between the causal frames and the Landau frame. 

The frame change to Landau frame is straightforwadly obtained from (\ref{eq:tmunu11})
 \begin{subequations}
\begin{align}
\epsilon_{\text{Landau}}&=\epsilon + 2\,a_2\,\eta  \left(\frac{2}{3}\frac{\dot{\epsilon}}{\epsilon}+\nabla \cdot u \right) \,, \label{changetoLandau0a}\\
u^\mu_{\text{Landau}} &= u^\mu +\frac{2\, a_1\,\eta }{3\epsilon}\left(\dot{u}^{\mu}+\frac{1}{3} \frac{\nabla_{\perp}^{\mu} \epsilon}{\epsilon}\right) \,.
\end{align}
\label{changetoLandau0}
 \end{subequations}

In our simulations we use the transport coefficients of the microscopic holographic theory that will be presented in the next Section  \cite{,Herzog:2002fn, Natsuume:2007ty,VanRaamsdonk:2008fp}:
\begin{subequations}
\begin{eqnarray}
\eta &=& \frac{s}{4\pi} ~,  \\
\tau_{\pi} &=& \frac{1}{24 \pi T}\left(18-9\ln 3+\sqrt{3}\pi \right)  ~, \\
\lambda_1 &=& \frac{3s}{32\pi^2 T}   ~, \\
\lambda_2 &=& \frac{2 \pi T}{27} \left( \sqrt{3}\pi -9\ln 3 \right)   ~, \\
\lambda_3 &=& 0   ~.
\end{eqnarray}
\end{subequations}

For details on the implementation of the hydrodynamic equations in the numerical codes and convergence tests see Appendix A.

%%%%%%%%%%%%%%%%%%%%%%%%%%%%%%%%%%%%%%%%%%%%%%%%

\section{Holography: microscopic evolution}
\label{holography}
%%%%%%%%%%%%%%%%%%%%%%%%%%%%%%%%%%%%%%%%%%%%%%%%
%Holographic model and initial data. Numerical relativity.
Our model is Einstein's gravity with negative cosmological constant coupled to a massless scalar field in four spacetime dimensions.\footnote{The scalar field is used to form a black brane by prompt collapse of the initial data. This prompt colapse happens immediately after initialization of the code and this timescale is well separated from the timescales relevant for the physics studied in this paper. Thus, after the immediate collapse the scalar field is completely negligible and the evolution will be well described by pure gravity. This pure gravity theory can be obtained as a consistent truncation in a set of top-down theories which includes the ABJM theory \cite{Aharony:2008ug}.}
By holography, it describes the decoupled sector of the stress tensor on the conformal field theory side.
The real-time quantum dynamics on the field theory maps to the bulk classical dynamics of Einstein's gravity in anti-de Sitter (AdS) space coupled to a massless scalar field. We use numerical relativity techniques to solve for the classical dynamics of the bulk fields employing the same code as in \cite{Bantilan:2020pay,Bantilan:2020xas}. For more details of our implementation see these two references and Appendix B. 

We consider the gravitational collapse of massless scalar field in the 3+1 dimensional Poincar\'e patch of AdS using time symmetric initial data. The scalar field has an initial deformed Gaussian profile along the boundary directions and is localized in the AdS radial direction. The initial data is ``strong'' in the sense that there is a trapped surface on the initial Cauchy surface. Therefore, one can think of our initial data as corresponding to a highly deformed black brane. 

\begin{figure}[t]
	\centering	\includegraphics[width=0.64\textwidth]{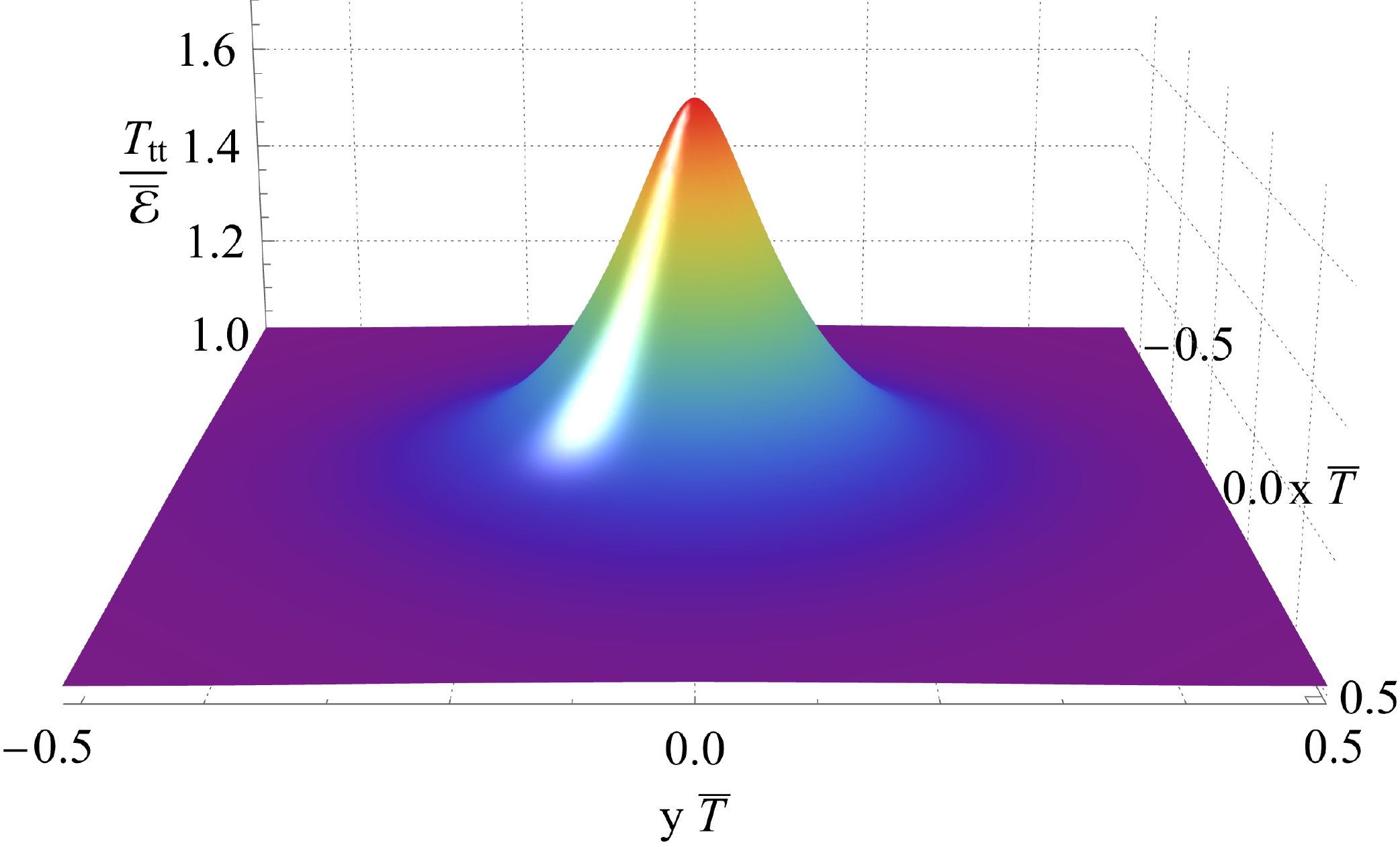}	
	\includegraphics[width=0.64\textwidth]{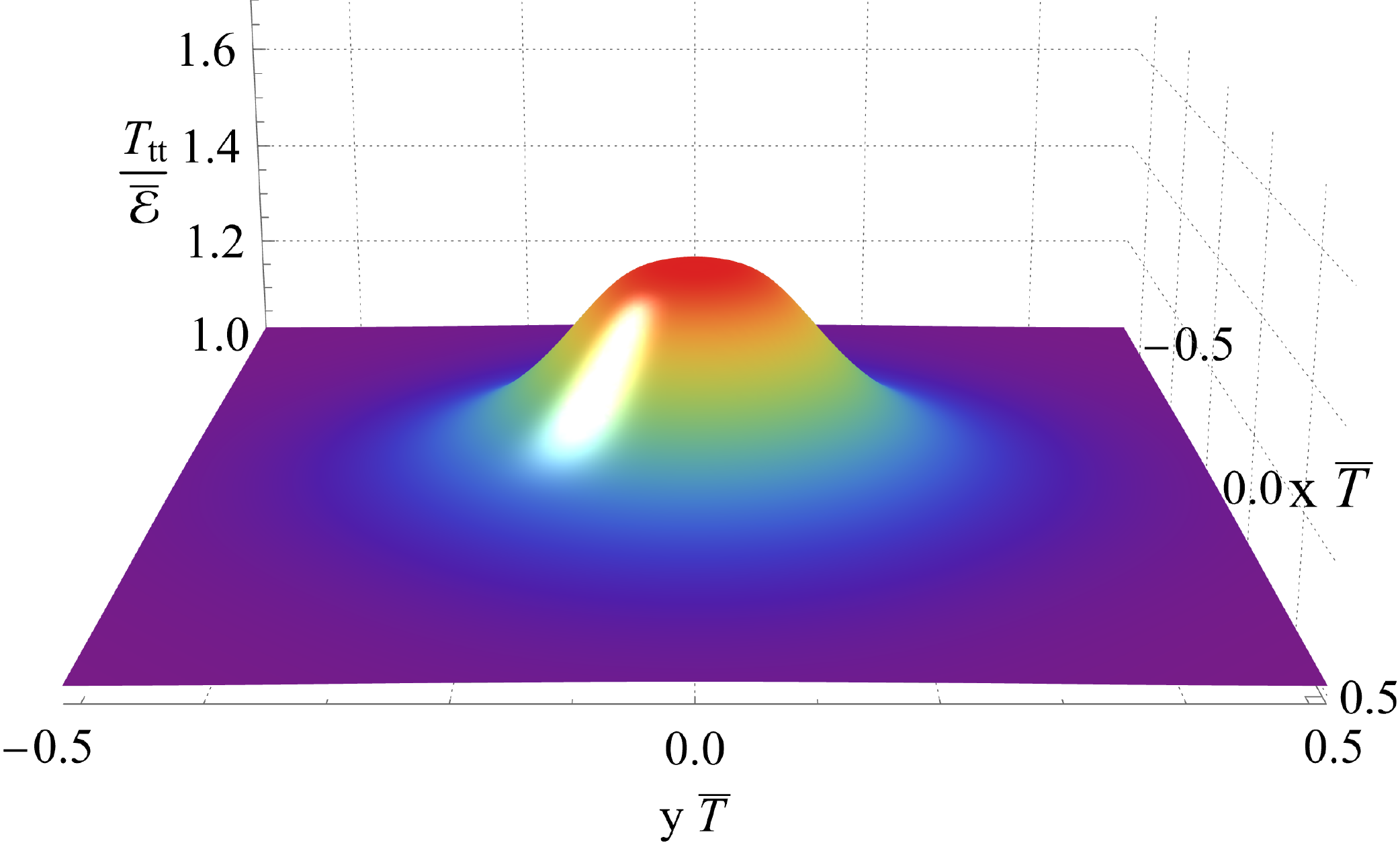}
	\includegraphics[width=0.64\textwidth]{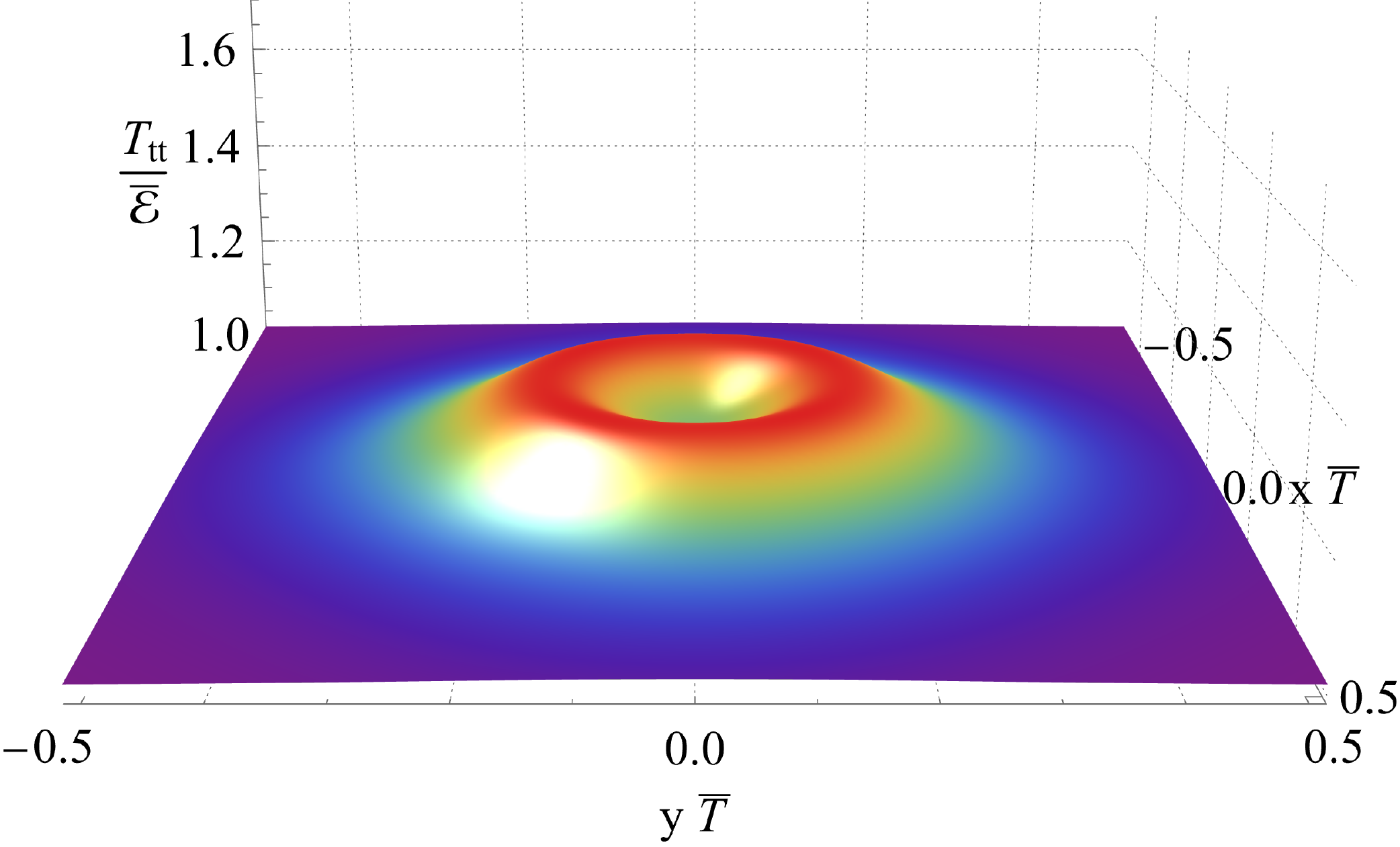}
	\caption{Snapshots of the energy density in the lab frame, $T_{tt}$, of the holographic solution at times $t\overline{T}=0, 0.08, 0.16$, from top to bottom.}
	\label{Holographic_energy_lab_frame}
\end{figure}

From the dual CFT point of view, the  initial state corresponds to a large localized perturbation on top of a homogeneous plasma at temperature $\overline{T}$. The initial energy density profile on the field theory is shown in Fig. \ref{Holographic_energy_lab_frame} (top). The initial data is rotationally symmetric and has vanishing initial velocities (since the initial data is time-symmetric), even though our code does not assume rotational symmetry along the boundary directions. We have also considered initial data with small deformations that break the rotational symmetry along the boundary directions. However, the non-rotationally symmetric modes decay exponentially and by the time the system approaches the hydrodynamic regime,  those modes have already decayed. We measure all quantities in units of $\overline{T}$ or $\overline{\mathcal E} \equiv \frac{2}{3} A \overline{T}^3$ and we use Cartesian coordinates $\{t,x,y\}$ to label the boundary directions.

The time evolution of the system is shown in Fig. \ref{Holographic_energy_lab_frame} in three representative snapshots at times $t\overline{T}= 0,0.08,0.16$. The initial peak explodes and the system expands and disperses away, and at late times it relaxes to a homogeneous solution. 

%%%%%%%%%%%%%%%%%%%%%%%%%%%%%%%%%%%%%%%%%%%%%%%%

\section{Hydrodynamics: evolutions}
\label{hydro}
%%%%%%%%%%%%%%%%%%%%%%%%%%%%%%%%%%%%%%%%%%%%%%%%
%Hydrodynamics: constitutive relations
In the following we study the applicability of hydrodynamics in this system. We do this in two ways: first we check the constitutive relations pointwise in spacetime and, second, we perform time evolution of the hydrodynamic equations using the holographic solution as  initial data.

We start by evaluating the constitutive relations of hydrodynamics \eqref{constitutive0sheartensor0} pointwise in spacetime using the holographic data to calculate the various terms in this expression. 
 In the Landau frame, gradient corrections in the hydrodynamic expansion are transverse to the velocity and, together with conformal symmetry, this  implies that they vanish at the center of a spherically symmetric system. Away from the center however the gradients are non-trivial; we choose $\{x,y\}\overline{T}\simeq \{0.12,0\}$ as a representative point where to evaluate the constitutive relations. 

\begin{figure}[t]
	\centering	\includegraphics[width=0.64\textwidth]{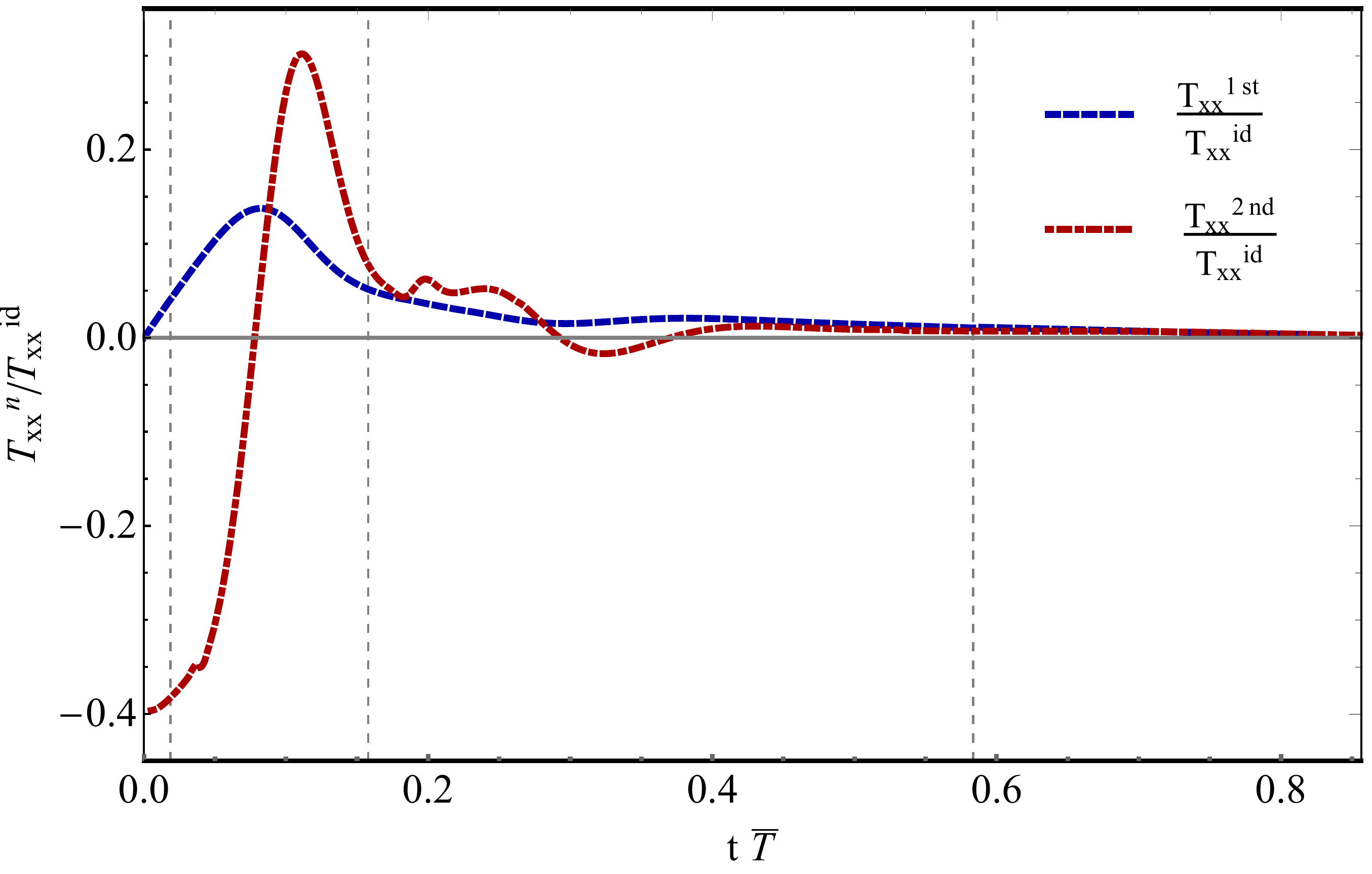}	
	\caption{Ratios $T_{xx}^{n}/T_{xx}^{ideal}$, $n=1,2$, of the constitutive relations (\ref{constitutive0sheartensor0}) evaluated at $\{x,y\}\overline{T}\simeq \{0.12,0\}$ as a function of time for the holographic solution. The vertical dashed lines show the times at which we initialize the hydrodynamics codes $t\overline{T}\simeq 0.019, 0.16, 0.58$. }
	\label{Constitutive_relations_Frame2}
\end{figure}

In Fig.  \ref{Constitutive_relations_Frame2} we compare the size of the 1st and 2nd order corrections to the ideal term in the derivative expansion of the stress tensor by plotting  $T_{xx}^{1st}/T_{xx}^{id}$ and $T_{xx}^{2nd}/T_{xx}^{id}$ as functions of time. This figure indicates that, according to the constitutive relations, the system is initially far from equilibrium, as second order gradients are as large as $40\%$ compared to the ideal terms. The size of the derivative corrections quickly decays and the system has hydrodynamized at times around $t\overline{T}\simeq 0.5$, if our criterium is that the ratios $T_{xx}^{1st}/T_{xx}^{id}$ and $T_{xx}^{2nd}/T_{xx}^{id}$ are smaller than $2\%$.

%Hydrodynamic evolutions
We now consider the evolutions of the hydrodynamic equations of the various theories using initial data from the holographic  solution. We start by diagonalizing the holographic stress tensor and obtaining the local energy density and velocities (these are the energy and velocities in the Landau frame), which will be used to initialize the hydro codes. In the particular case of BRSSS, we have to also specify the dissipative part of the stress tensor $\Pi^{\mu\nu}$ from the microscopic holographic data. But notice that
the use of time symmetric initial data implies that at $t=0$ the dissipative part of the stress tensor vanishes identically and so in the subsequent evolution it is fairly small and hence one could initialize the code using {very small initial data for the viscous tensor $\Pi^{\mu\nu}$. However, initially the second order terms are large and therefore, alternatively,  one could initialize de code using the dissipative part of the stress tensor computed with the first and second order consitutive relations \eqref{constitutive0sheartensor0}, which is not small. These two possible ways of initializing the code match in the hydrodynamic regime (up to 3rd order terms), but under these far-from-equilibrium conditions they differ. We choose the second option, which captures the initial presence of large gradients.

For the evolutions in the BDNK theory,  we consider the following causal frames: 
\begin{subequations}
	\begin{align}
	\text{Frame} \ 1: & \quad\quad\quad a_1=10,\quad  a_2=10 \, , \\
	\text{Frame} \ 2: & \quad\quad\quad a_1=6,\quad \, \  a_2=4 \, ,
	\end{align}
	\label{frames}
\end{subequations}
which satisfy the hyperbolicity conditions \eqref{gap}.
We obtain the energy and velocities in the causal frame from the Landau frame by inverting the expressions \eqref{changetoLandau0} and neglecting second order terms. The inverted expressions are the following: 
 \begin{subequations}
	\begin{align}
		\epsilon&=\epsilon_{\text{Landau}} - 2\,a_2\,\eta  \left(\frac{2}{3}\frac{\dot{\epsilon}_{\text{Landau}}}{\epsilon_{\text{Landau}}}+\nabla \cdot u_{\text{Landau}} \right) \,, \label{changetoLandau0a}\\
		u^\mu &= u^\mu_{\text{Landau}} -\frac{2\, a_1\,\eta }{3\epsilon}\left(\dot{u}_{\text{Landau}}^{\mu}+\frac{1}{3} \frac{\nabla_{\perp}^{\mu} \epsilon_{\text{Landau}}}{\epsilon_{\text{Landau}}}\right) \,,
	\end{align}
	\label{changetoLandau0_inverted}
\end{subequations}

Where we have used the fact that in the first order terms we can replace $\epsilon=\epsilon_{\text{Landau}}$, $u^\mu = u^\mu_{\text{Landau}}$ as the difference is second order.
Recall that the evolution equations of BDNK are of second order and hence one also has to specify the time derivatives of the evolved variables in the initial data. We compute these by using the holographic data to obtain the constitutive relations in the corresponding causal frame as functions of time using (\ref{changetoLandau0_inverted}), and then calculating the time derivatives in the causal frame. Other procedures might be equivalent in the hydrodynamic regime up to second order terms, but in the far from equilibrium regime they will generically differ. 

We start the discussion by initializing the hydrodynamic codes using the holographic data at $t\overline{T}\simeq 0.019$, when the system is still far from equilibrium according to the constitutive relations. Fig. \ref{fig:off_center_evolutions} (top) shows the evolution of the energy density in the lab frame at a representative off-center point of the domain $\{x,y\}\overline{T}=\{0.17,0\}$ obtained using ideal (dashed green), BRSSS (dashed blue), BDNK in frame 2 (purple curves) and the holographic solution (solid black curve). 
For the BDNK evolutions,  we plot two different quatities: the result of obtaining $T_{tt}$ from the evolved variables using \eqref{eq:tmunu11} (solid line), and the result of evaluating only the 0th order terms (ideal tems) in \eqref{eq:tmunu11} (dotted line). The reason for doing this is to show explicitly the size of the first order terms compared to the ideal terms.
 As this figure shows, the three hydrodynamic theories that we consider initialized at $t\overline{T}\simeq 0.019$ exhibit a similar level of (large) disagreement with the microscopic theory throughout the evolution. The fact that none of the theories of hydrodynamics provides an accurate description of the evolution of the fluid confirms that at  this initial time, the system is still very far from the hydrodynamic regime.  This is expected since the gradients are large initially and each theory has a different UV completion. Therefore, there is no reason why these theories should agree with each other away from the hydrodynamic regime.

\begin{figure}[t!]
	\centering	
	\includegraphics[width=0.64\textwidth]{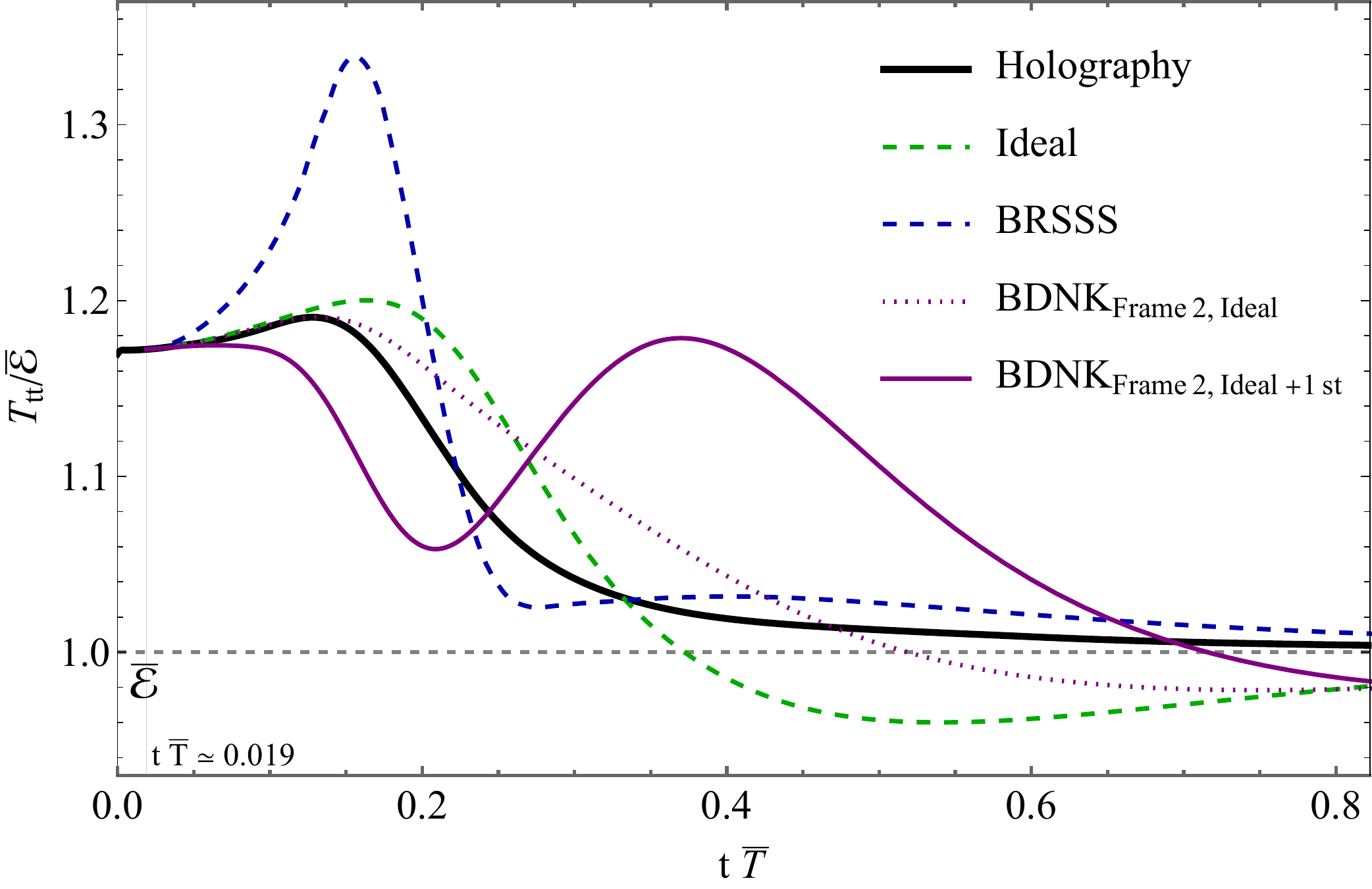}	\includegraphics[width=0.64\textwidth]{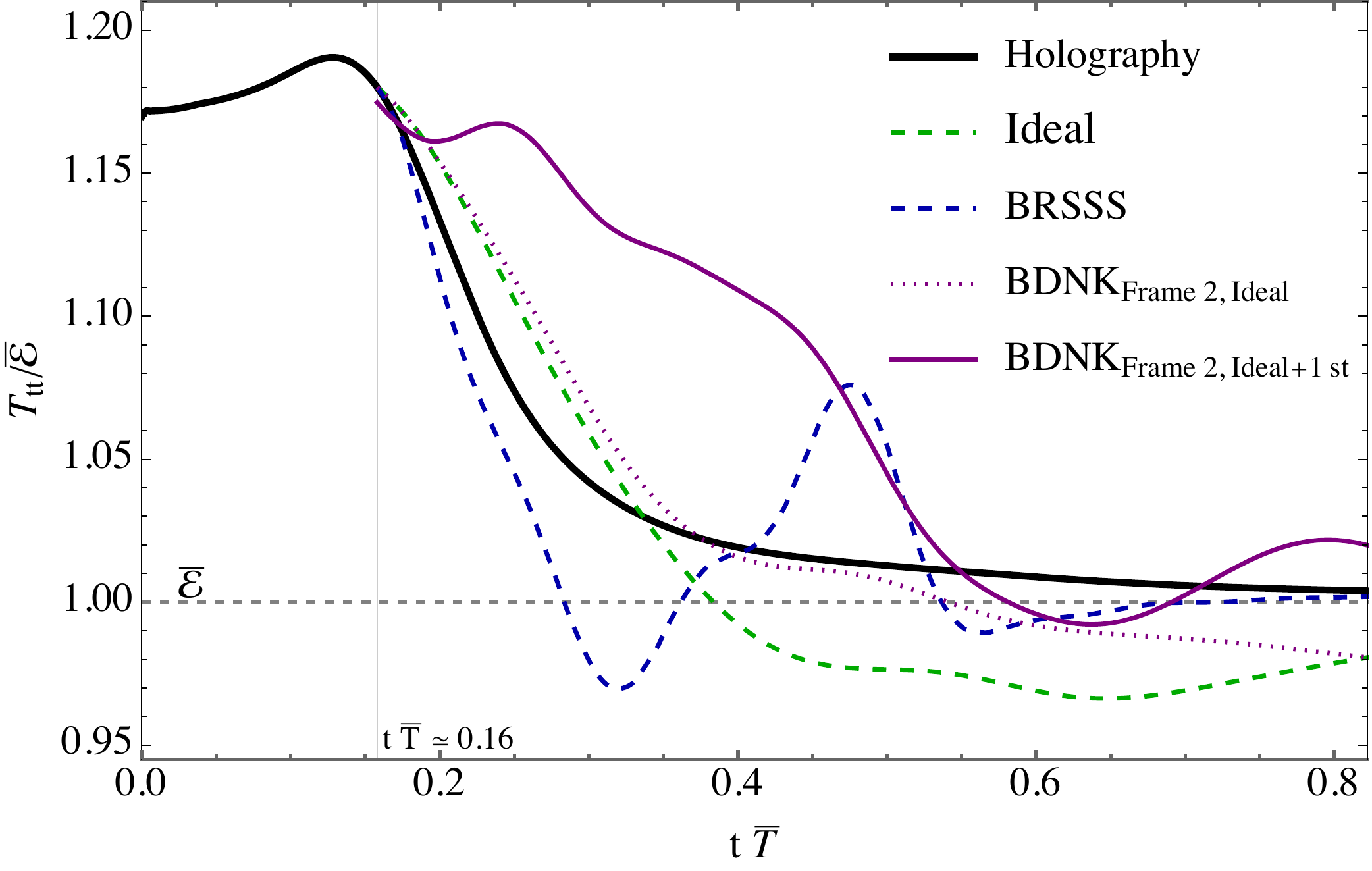}
	\includegraphics[width=0.64\textwidth]{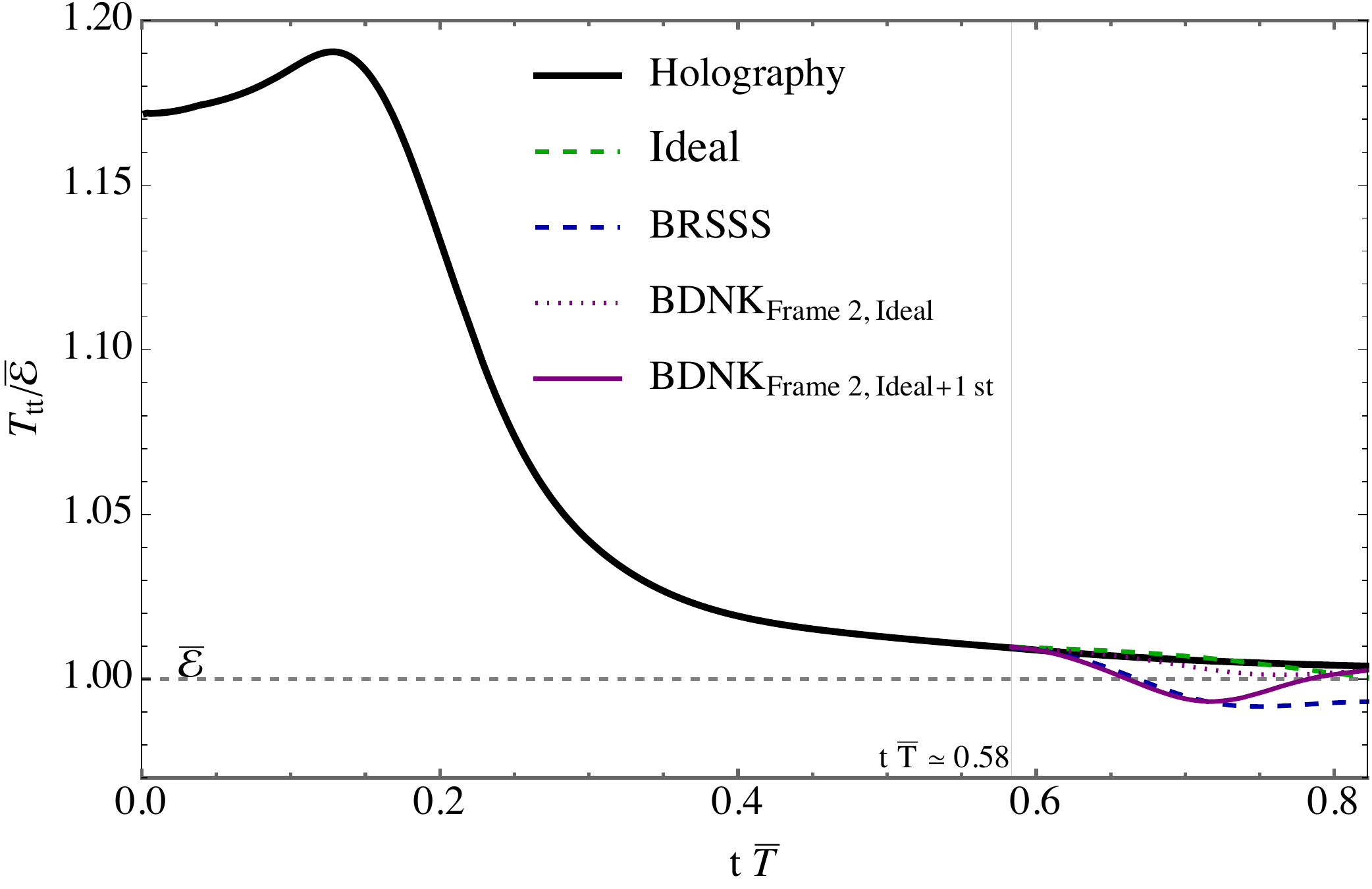}
	\caption{Energy density in the lab frame, $T_{tt}$, at  an off center location  $\{x,y\}\overline{T}=\{0.17,0\} $ as a function of time. Black continuous line corresponds to the microscopic (holographic) solution. We include the results of the hydrodynamic evolutions initialised with holographic data at $t\overline{T}\simeq 0.019, 0.16, 0.58$, from top to bottom. For the BDNK evolutions, we also include the result of using only the 0th order terms (the ideal part) in (\ref{eq:tmunu11}), in dotted lines. The other components of the stress energy tensor behave in a quantitatively similar manner.}
	\label{fig:off_center_evolutions}
\end{figure}

We continue the discussion by considering hydrodynamic evolutions with holographic initial data at a later time $t\overline{T} \simeq 0.16$, when the size of the gradients is smaller but the system has not hydrodynamized yet according to the constitutive relations, see Fig. \ref{fig:off_center_evolutions} (middle panel). Indeed, at this initial time the constitutive relations indicate that the first and second order terms are comparable, see the second vertical line in Fig. \ref{Constitutive_relations_Frame2}. We find that all hydrodynamic evolutions still exhibit large deviations from the microscopic theory which confirms that the system has not reached the hydrodynamic regime yet. Again, the deviations of BRSSS and BDNK from the microscopic theory are comparable.

Finally, we consider hydrodynamic evolutions with holographic initial data at a later time $t\overline{T} \simeq 0.58$, when the size of the gradients is small, see Fig. \ref{fig:off_center_evolutions} (bottom panel).
At $t\overline{T} \simeq 0.58$, by checking the constitutive relations \eqref{constitutive0sheartensor0} we find that the first order gradients have a maximum value in the whole domain of the order of $~1\%$, i.e., $T_{xx}^{1st}/T_{xx}^{id}\sim 1\%$, and similarly for the second order gradients, i.e., $T_{xx}^{2nd}/T_{xx}^{id}\sim 1\%$. These values are small, and we expect the hydrodynamic evolutions to follow the microscopic solution when initialized at this time. Our results shown in the bottom panel of Fig. \ref{fig:off_center_evolutions} confirm this expectation. This is true within the numerical errors of the holographic solution, which are inherited by the hydrodynamic evolutions through the initial data.  We have checked that the energy density in the lab frame for the hydrodynamic evolutions differs by less than $\sim 2\%$ for BRSSS, BDNK and  ideal hydrodynamics, compared to the microscopic evolution in whole the domain and at all times beyond $t\overline{T} \simeq 0.58$. 
These results also confirm that the various theories of hydrodynamics provide equivalent descriptions of the system, further supporting the applicability of hydrodynamics. 

Fig. \ref{fig:off_center_evolutions} (bottom) suggests that ideal hydrodynamics provides a better description of the system at late times than viscous hydrodynamics. The reason might be the following.  Our particular choice of initial data implies that the holographic system at $t=0$ is exactly described by the constitutive relations of ideal hydrodynamics \eqref{constitutive0} with $\Pi^{\mu\nu}=0$   (even if second order terms are large and the system is not within the regime of hydrodynamics) \footnote{The argument of why the system at t=0 is exactly described by the constitutive relations of ideal hydrodynamics is as follows. Rotational symmetry and vanishing of the initial velocities imply that the stress tensor is diagonal (in cartesian coordinates) at t=0, and that $T_{xx}=T_{yy}$. Conformal symmetry implies that the trace is vanishing, $-T_{tt}+T_{xx}+T_{yy}=0$, obtaining $T_{xx}=T_{tt}/2$, which is the equation of state (\ref{eos_conformal}). Thus, eq. \eqref{constitutive0} with $\Pi^{\mu\nu}=0$ exactly describes the microscopic holographic tensor at t=0.  }. This aspect may partially survive through the far from equilibrium region explaining why ideal hydrodynamics provides a better description of the microscopic system than viscous hydrodynamics around the hydrodynamization time. We think that this might be a particularity of the specific system under consideration, and we wonder if this could be a generic feature of systems sharing similar initial conditions.

See Appendix C for more details and further examples.

%%%%%%%%%%%%%%%%%%%%%%%%%%%%%%%%%%%%%%%%%%%%%%%%

\section{Discussion}
%\label{Discussion}
%%%%%%%%%%%%%%%%%%%%%%%%%%%%%%%%%%%%%%%%%%%%%%%%
We have used holography to obtain, from first principles, the real-time quantum dynamics of a large-$N$ strongly coupled conformal field theory initially far from equilibrium which relaxes to a hydrodynamic regime. This constitutes a microscopic solution that we used to test the applicability of hydrodynamics. This solution is a very simple toy model that captures one important aspect of the physics of the QGP created in heavy-ion collision experiments: the initial far-from-equilibrium conditions and subsequent relaxation to a hydrodynamic regime. 

We have considered the applicability of three theories of hydrodynamics, namely  ideal hydrodynamics, BRSSS and the newly formulated BDNK theory. For each of these theories, we evolve the equations of motion numerically using the microscopic solution as initial data at different times to explore the non-linear and far-from-equilibrium regimes. 
We assess the applicability of hydrodynamics by comparing both the constitutive relations, as previously done in the literature, and the evolution of the fluid as predicted by each theory. These comparisons are shown in Fig.  \ref{fig:off_center_evolutions}, and we outline the main conclusions in the following.

%1
The predictions of the different hydrodynamic theories differ from the microscopic solution and from each other in the far from equilibrium region, $t\overline{T}\simeq  0.019,0.16$ in our case.  This is expected since each theory has a different a UV completion and, in this regime, the latter is relevant. 
%2
On the other hand, at late times, $t\overline{T}\simeq 0.58$ in our case, the gradients are small and hydrodynamics provides a good description of the system in the sense that the three theories that we have considered agree well with the microscopic evolution and with each other.

%3
In this article we have studied BDNK as a theory of relativistic viscous hydrodynamics having in mind applications to the QGP. BDNK is particularly interesting because its only evolution variables are the thermodynamic quantities such as energy density and fluid velocities. To initialize the BDNK equations, we have to consider data for the fundamental thermodynamic variables and their first time derivatives in the causal frame obtained by inverting \eqref{changetoLandau0} and neglecting higher order terms. This is only justified in the hydrodynamic regime. 
%4
We  performed evolutions in different causal frames; only when the system is in the hydrodynamic regime we find that  the physics is not affected by the choice of causal frame, up to second order terms. Away from the hydrodynamic regime,  evolutions carried out in different frames in general differ. 
 
%5
In the QGP created in heavy-ion collisions the hydrodynamic evolutions provide a good description of the system even if gradients are not very small,  which sometimes is known as 'unreasonable success of hydrodynamics'. In our case, we do not find such an unreasonable success, but instead we found that hydrodynamics applies precisely when it should. The theories that we considered and the initial data are very different from the QGP case, so there is a priori no reason why the success should happen also in our scenario.

%6
One may wonder which causal formulation of viscous hydrodynamics, BRSSS or BDNK, provides a better description of the system. However, this may not be the right question to ask given that all theories of hydrodynamics should be equivalent in the regime of validity of hydrodynamics. Therefore, the question  that one would really like to answer is which of these theories  (if any!) provides a better (i.e., more in accordance to the microscopic theory) description of the system slightly outside the regime of hydrodynamics. It is likely that the answer depends on the details of the system that one wants to model and the initial conditions. In our case, where we have full control of the microscopic theory, we did not find that either BRSSS or BDNK provides a better description of the evolution outside the hydrodynamic regime. More studies are needed to address this question. 

%Definition of hydrodynamics as evolutive theories
Hydrodynamics can be defined by considering a specific well-posed theory subject to suitable initial conditions.  We may wonder if the constitutive relations evaluated pointwise in spacetime on the solution obtained by solving the initial value problem in hydrodynamics and those obtained from the microscopic solution may provide a different answer regarding the regime of applicability of hydrodynamics. This is an important question since in practical applications in general one does not have a theoretical control of the microscopic theory, and
our solutions provide a concrete example where this question can be addressed. Above we have defined the hydrodynamization time using the second approach, obtaining $t\overline{T} \simeq 0.5$. Alternatively, we could define the hydrodynamization time using the first approach, defined as the time at which, if we initialize the hydrodynamic code with microscopic data, the stress tensors along the rest of the evolution differ less than a $2\%$ with the holographic one. For this case we obtain hydrodynamization times compatible with $t\overline{T} \simeq 0.5$. Thus, both approaches provide a compatible answer. Having two different theories of relativistic viscous hydrodynamics, i.e., BDNK and BRSSS, allows us to use a third criterium, 
%which could be implemented 
which might be useful
in practical applications where the microscopic theory cannot be solved. If initialising both theories at a given time using suitable initial conditions, the evolutions differ by less than some prescribed small amount at all times, then the system is in the hydrodynamic regime. The caveat is that one has to be careful to ensure that the initial data for both theories is the ``same''.
 
 %Also, we would like to explore a third criterium, which might be useful in practical applications when the microscopic theory cannot be solved. 
 %Having two different theories of relativistic viscous hydrodynamics, i.e., BDNK and BRSSS, we can initialise both theories at a given time using suitable initial conditions, and if the evolutions differ by less than some prescribed small amount at all times, then the system is in the hydrodynamic regime. The caveat is that one has to be careful to ensure that the initial data for both theories is the ``same''.

%Future
With this work we initiate a program to study evolutions of the BDNK equations and to test the applicability of causal viscous hydrodynamic theories by comparing with microscopic holographic evolutions. Possible extensions include non-conformal and charged theories and initial data that models heavy-ion collisions. We hope that our work, together with \cite{Pandya:2021ief}, provide the first steps towards the implementation of the BDNK equations to describe relevant physical systems like the QGP or neutron star mergers. 

%%%%%%%%%%%%%%%%%%%%%%%%%%%%%%%%%%%%%%%%%%%%%%%%

\subsection*{Acknowledgements}
%\label{Acknowledgements}
%%%%%%%%%%%%%%%%%%%%%%%%%%%%%%%%%%%%%%%%%%%%%%%%
We thank Anton Faedo and Jorge Noronha for useful discussions.
Y.B. and P.F. are financially supported by the European Research Council grant ERC-2014-StG 639022-NewNGR. 
Y.B. is also supported by the Academy of Finland grant no. 333609.
P.F. is also supported by a Royal Society University Research Fellowship (Grant No. UF140319 and URF\textbackslash R\textbackslash 201026). 
We thank the MareNostrum supercomputer at the BSC (activity Id FI-2020-1-0007) for significant computational resources.

\begin{appendix}
	
%%%%%%%%%%%%%%%%%%%%%%%%%%%%%%%%%%%%%%%%%%%%%%%%%%%%%%%%%
\section{Implementation of hydrodynamic codes and tests}
\label{Aappedix}
%%%%%%%%%%%%%%%%%%%%%%%%%%%%%%%%%%%%%%%%%%%%%%%%%%%%%%%%%
In our hydrodynamic codes we work directly with the primitive variables, namely energy density and velocities of the fluid, and the viscosity tensor in the case of BRSSS. This implies that our code cannot deal with shocks. That said, we point out that for the class of initial conditions that we have considered in this article, we did not observe the formation of shocks or steep features in the fluid flows that would require the use high resolution shock capturing techniques.   

For the BRSSS case, we follow \cite{Green:2013zba} and we impose the constraints $u^\mu u_\mu=-1$, $u^\mu \Pi_{\mu\nu}=0$ and $\Pi^\mu_{\phantom{\mu}\mu}=0$ algebraically to reduce the number of dynamical variables to 5, which we take to be $\U=\{\rho,u_x,u_y,\Pi_{xx}, \Pi_{xy}\}$. Then we use {\tt Mathematica} to solve the hydrodynamics equations of motion in terms of the time derivatives of $\U$ and write the them as
\begin{equation}
\partial_t \U = \mathcal{F}(\U,\partial_i\U)\,.
\label{eq:eoms_brsss_sym}
\end{equation}
The ideal case can be recovered from the BRSSS case by setting to zero all the transport coefficients and $\Pi_{xx}=\Pi_{xy}=0$, as well as their derivatives.  In our code, we discretize the spatial derivatives in \eqref{eq:eoms_brsss_sym} using 6th order finite difference stencils and  we integrate them forward in time using a standard RK4 time-integrator.  For simplicity, we imposed periodic boundary conditions; therefore, in our hydrodynamic simulations we have to choose a large enough domain to avoid boundary effects during the duration of the simulations. In practice this is not a problem since the hydrodynamic simulations are very cheap. 

The equations of motion for BDNK are second order in time and space. In this case, we implement the constraint $u^\mu u_\mu=-1$ algebraically to reduce the number of independent variables to 3; we choose $\U=\{\rho,u_x,u_y\}$. Again, we use {\tt Mathematica} to solve for the second time derivatives of the dynamical variables, and re-write the equations of motion as
\begin{equation}
\partial_t^2 \U = \mathcal{G}(\U,\partial_t\U,\partial_i\U,\partial^2_{t,i}\U,\partial^2_{i,j}\U)\,.
\label{eq:eoms_bdnk_sym}
\end{equation}
One could write \eqref{eq:eoms_bdnk_sym} as a first order in time system of equations in the obvious way, e.g., $\mathcal{P}\equiv\partial_t \U$, and use a standard integrator such as RK4. However, we found that the resulting system was numerically unstable; it is not clear to us what is the origin of this instability. We did not attempt to write \eqref{eq:eoms_bdnk_sym} as a fully first order system by further defining $\mathcal{V}_i \equiv \partial_i\U$.

As in the BRSSS and Ideal cases, in BDNK we discretise the spatial derivatives using 6th order finite differences and we impose periodic boundary conditions.  To proceed, we implemented two different time integrators. 

First, we implemented an implicit second order in time scheme by discretizing the time derivatives of a given variable $f$ at the time $t^n=n\,\Delta t$ as:
\begin{equation}
\begin{aligned}
\partial_t\U& \to \frac{\U^{n+1}_i-\U^{n-1}_i}{2\,\Delta t}\,,\\
\partial^2_{t}\U& \to \frac{\U^{n+1}_i-2\,\U^n_i+\U^{n-1}_i}{\Delta t^2}\,,
\end{aligned}
\end{equation} 
where $i$ denotes a collection of indices that labels a given spatial grid point. In this way, for known $\U^n_i$ and $\U^{n-1}_i$, the discrete equations of motion at time $t^n$ become a non-linear algebraic system for the values of the variables at the next time level, $\U^{n+1}_i$, which can be solved using standard techniques; in our case, for the BDNK equations we used a Newton-Raphson algorithm.  This method is robust and it works well in practice, and it has been successfully used in simulations of black hole binary mergers \cite{Pretorius:2004jg} and  in our holographic simulations \cite{Bantilan:2012vu,Bantilan:2020pay,Bantilan:2020xas}. However, for high spatial resolutions, it becomes considerably slower and memory-demanding than an explicit time integrator such as RK4. Therefore, for practical applications that one could easily run on a laptop, such as the ones presented here, it would be desirable to have a stable explicit time integrator for BDNK. We now turn to this. 

We have successfully implemented an explicit time integrator for the BDNK equations of motion, written as second order in time, i.e., \eqref{eq:eoms_bdnk_sym}. The analogues of the Runge-Kutta methods for second order equations are known as Runge-Kutta-Nystr\"om Generalized (RKNG) methods. We will review them here since they may not be as well-known as the standard first order  methods. Here we follow \cite{Fine:1987ul} since we have implemented their RKNG34 scheme. This method is competitive in terms of both accuracy and efficiency with respect to RK methods of similar orders, see \cite{Fine:1987ul} for detailed comparisons. 

Consider a system of $N$ second order ordinary differential equations:
\begin{equation}
y''(x) = f(x,y(x),y'(x))\,,
\end{equation}
on an interval $I = [x_0,x_F]$ with initial conditions
\begin{equation}
y(x_0)=y_0\,,\quad y'(x_0)=y'_0\,.
\end{equation}
The generalization of this method to PDEs is straightforward, just as in the standard RK4 case. An explicit RKNG method of $s$ stages allows to compute the approximations  $y_{n+1}$ and $y_{n+1}'$ of the solution $y(x_{n+1})$ and its derivative $y'(x_{n+1})$ at $x_{n+1}\in I$ from their values in the previous steps, $y_{n}$ and $y_{n}'$, as follows: 
\begin{align}
y_{n+1} &= y_n + h\,y_n' + h^2\sum_{i=1}^s b_i\,f_i\,, \label{eq:RKNG1}\\
y'_{n+1}& = y'_n + h\sum_{i=1}^s b'_i\,f_i\,, \label{eq:RKNG2}
\end{align}
where
\begin{align}
f_1 &= f(x_n,y_n,y'_n)\,,  \label{eq:RKNG3}\\
f_i & = f\bigg(x_n+c_i\,h,\,y_n+c_i\,h\,y'_n+h^2\sum_{j=1}^{i-1}a_{ij}\,f_j,\nonumber\\
&\hspace{1cm}y_n'+h\sum_{j=1}^{i-1}a'_{ij}f_j\bigg)\,,\quad i=2,\ldots,s\,,  \label{eq:RKNG4}
\end{align}
and $h=x_{n+1}-x_n$. Here $c_i$, $b_i$ and $b_i'$ are $s$-dimensional constant vectors, and $a_{ij}$ and $a'_{ij}$ are lower triangular $s\times s$ constant matrices; together, they constitute the parameters that define the method. An RKGN method is said to have order $p$ if the local convergence order for both the approximation $y_{n+1}$ that of the derivative $y'_{n+1}$ is $p$, 
\begin{align}
y_{n+1} - y(x_{n+1}) &= T_{p+1}\,h^{p+1} + O(h^{p+2})\,,\\
y'_{n+1} - y'(x_{n+1}) &= T'_{p+1}\,h^{p+1} + O(h^{p+2})\,,
\end{align}
where $T_{p+1}$ and $T'_{q+1}$ are the principal error functions of the solution and the derivative respectively and $h$ is the grid spacing. See \cite{Fine:1987ul} for more details. As in other multistep methods, such as standard Runge-Kutta methods, the global truncation error,  i.e., the accumulated error, is of order $O(h^p)$.

Ref. \cite{Fine:1987ul} gives a detailed derivation of a RKNG45 and a RKNG34 methods; here we only give the values of the parameters in \eqref{eq:RKNG1}--\eqref{eq:RKNG4} of the 5-stages RKNG34 scheme, which is a fourth order method, since this is what we implemented in our BDNK code as this method is competitive with the standard RK4 method. For the RKNG34 method of \cite{Fine:1987ul}, one has:
\begin{align}
b_i &= \left(\tfrac{19}{180},0,\tfrac{63}{200},\tfrac{16}{225},\tfrac{1}{120}\right)\,,\\
b'_i&= \left(\tfrac{1}{9},0,\tfrac{9}{20},\tfrac{16}{45},\tfrac{1}{12} \right)\,,\\
c_i&= \left(0,\tfrac{2}{9},\tfrac{1}{3},\tfrac{3}{4},1\right)\,,
\end{align}
with
\begin{align}
a_{21} &= \tfrac{2}{81}\,,\nonumber\\ 
a_{31} &= \tfrac{1}{36}\,,\,\, a_{32} = \tfrac{1}{36}\,,\\
a_{41} &= \tfrac{9}{128}\,,\,\, a_{42} = 0\,,\,\, a_{43} = \tfrac{27}{128}\,,\nonumber\\
a_{51} &= \tfrac{11}{60}\,,\,\, a_{52} = -\tfrac{3}{20}\,,\,\, a_{53} = \tfrac{9}{25}\,,\,\, a_{54} = \tfrac{8}{75}\,,\nonumber
\end{align}
and
\begin{align}
a'_{21} &= \tfrac{2}{9}\,,\nonumber\\
a'_{31} &= \tfrac{1}{12}\,,\,\, a'_{32} = \tfrac{1}{4}\,,\\
a'_{41} &= \tfrac{69}{128}\,,\,\, a'_{42} = -\tfrac{243}{128}\,,\,\, a'_{43} = \tfrac{135}{64}\,,\nonumber\\
a'_{51} &= -\tfrac{17}{12}\,,\,\, a'_{52} = -\tfrac{27}{4}\,,\,\, a'_{53} =-\tfrac{27}{5}\,,\,\, a'_{54} = \tfrac{16}{15}\,,\nonumber
\end{align}
and the remaining components of $a_{ij}$ and $a'_{ij}$ all vanish.

We study the convergence of our BDNK, BRSSS and ideal hydrodynamics codes by considering the evolution of a Gaussian initial profile for the energy density of the form,
\begin{equation}
\rho = \rho_0\bigg(1+\tfrac{A}{R\sqrt{2\pi}} \,e^{-\frac{(r-r_0)^2}{2R^2}}\bigg)\,,
\end{equation}
centred at $r=r_0$, with width $R$ and amplitude $A$, on top of a background energy density $\rho_0$. The initial values for all the remaining variables are chosen to be zero. For the runs shown below, we chose a computational domain of size $L=10$, $\rho_0=0.5$, $A=0.5$, $R=1$ and we centred the Gaussian at the centre of our grid.  For the BDNK simulations, we choose a frame with $a_1=a_2=10$. We monitor the convergence rate $Q_N(t)$,
\begin{equation}
Q_N(t) = \frac{1}{\ln 2}\,\ln\left(\frac{|| u_{4h}(t)-u_{2h}(t)||_1}{|| u_{2h}(t)-u_{h}(t)||_1}\right)\,,
\end{equation}
where $||\cdot ||_1$ denotes the numerical $L^1$-norm taken over our computational domain, and $h$ is the grid spacing. To compute these norms, we sum over all the evolution variables. 

\begin{figure}[h]
	\centering	
	\includegraphics[width=0.64\textwidth]{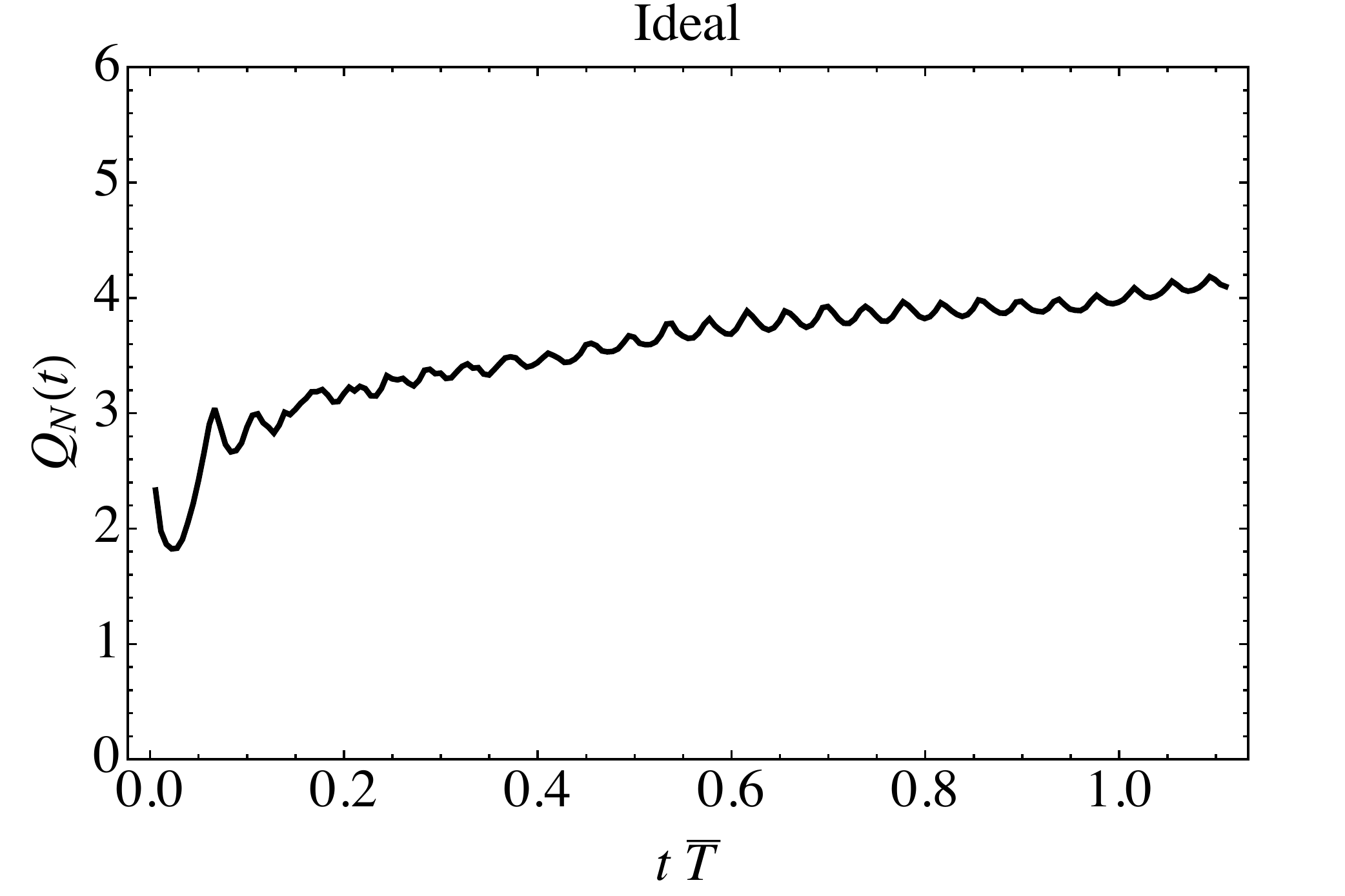}	
	\includegraphics[width=0.64\textwidth]{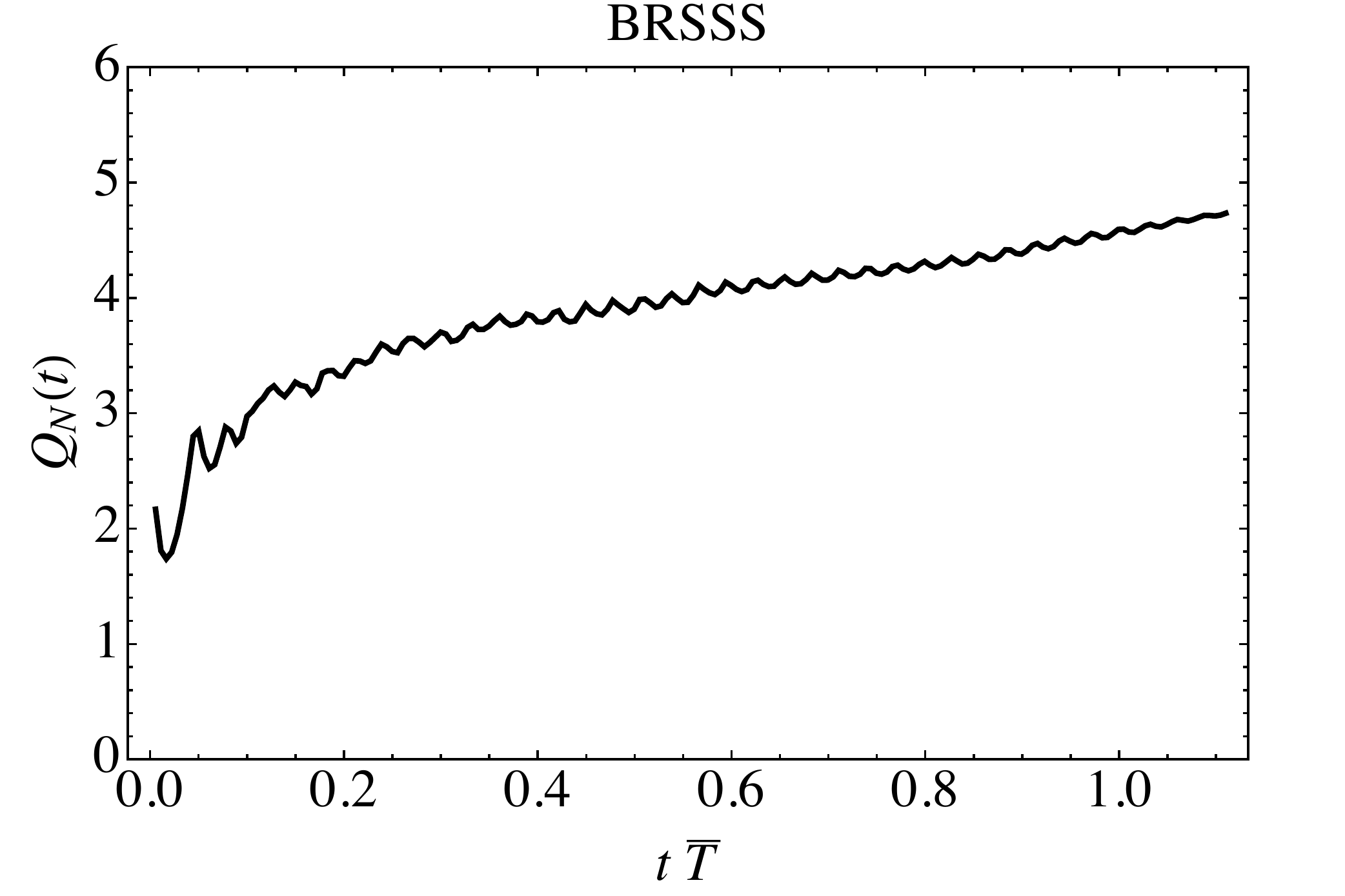}
	\includegraphics[width=0.64\textwidth]{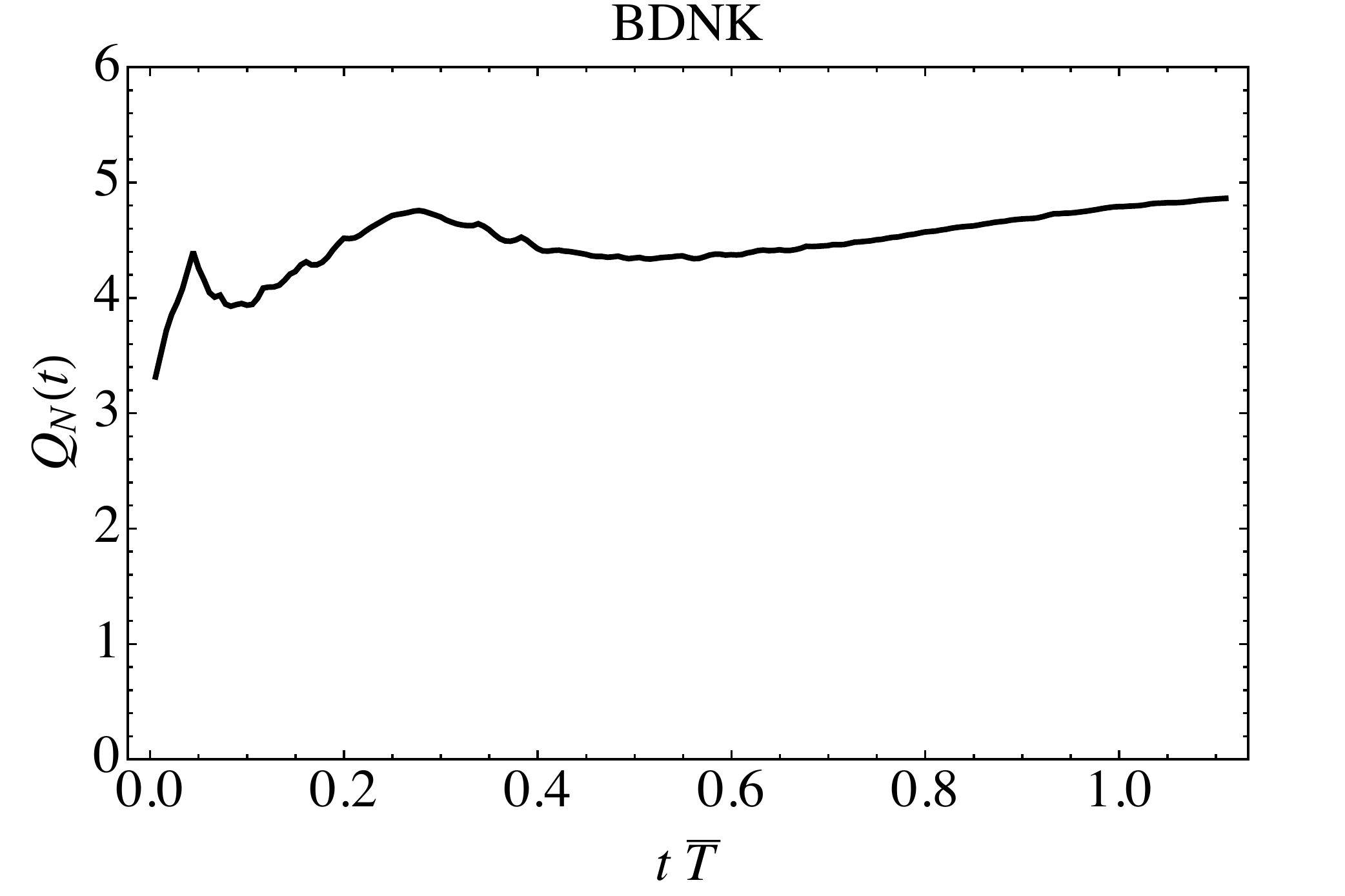}
	\caption{Convergence tests for our ideal (top), BRSSS (middle) and BDNK (bottom) codes. After an initial transient, the convergence rate approaches (and even surpasses) 4, as one would expect given our differencing and time integration schemes.  }
	\label{fig:conv_tests}
\end{figure}

The results of our convergence tests are displayed in Fig. \ref{fig:conv_tests} for the three theories that we consider in this article. As this figure shows, the convergence rate is close to 4 in all cases. Even though in our code we use 6th order spatial differences, this figure shows that the dominant error comes from the time integration. This justifies the use of 6th order Kreiss-Oliger dissipation.  Furthermore,  the convergence rate that we find is consistent with the order of the time integrators that we used in our code, namely RKNG34 for BDNK and RK4 for BRSSS and ideal hydrodynamics.

%%%%%%%%%%%%%%%%%%%%%%%%%%%%%%%%%%%%%%%%%%%%%%%%%%%%%%%%%
\section{Numerical relativity details}
\label{Cappedix}
%%%%%%%%%%%%%%%%%%%%%%%%%%%%%%%%%%%%%%%%%%%%%%%%%%%%%%%%%

In this appendix we summarize the basic aspects of our numerical relativity simulations. We use the same code as in \cite{Bantilan:2020pay,Bantilan:2020xas}, so more details can be found in these references. 

We solve the Einstein equations in AdS coupled to a massless scalar field in 3+1 dimensions using generalized harmonic coordinates, 
\begin{equation}
\begin{aligned}
&- \frac{1}{2} g^{\gamma \delta} g_{\alpha\beta, \gamma \delta} - 
{g^{\gamma\delta}}_{,(\alpha} g_{\beta) \gamma, \delta} - H_{(\alpha, \beta)} + H_\gamma {\Gamma^\gamma}_{\alpha\beta} \nonumber \\
&- {\Gamma^\gamma}_{\delta \alpha} {\Gamma^\delta}_{\gamma \beta} - \kappa \left( 2 n_{(\alpha} C_{\beta)} - (1+P) g_{\alpha\beta} n^\gamma 
C_\gamma \right) \nonumber \\
&=  \frac{2}{d-2} \Lambda g_{\alpha\beta} + 8\pi \left( T_{\alpha\beta} - 
\frac{1}{d-2} {T^\gamma}_\gamma g_{\alpha\beta} \right),
\end{aligned}
\end{equation}
where $n_\alpha=-\partial_\alpha t$ is the timelike, future-directed unit 1-form normal to slices of constant $t$, and $\kappa$ and $P$ are the damping parameters. Here $C^\alpha \equiv H^\alpha-\Box x^\alpha$ are the constraints and $H_\alpha$ are the source functions, which we can freely prescribe. We choose both the parameters and the source functions as in \cite{Bantilan:2020pay,Bantilan:2020xas}.  
 
For convenience, we couple gravity in AdS to a massless scalar field, since this allows us to easily construct far-from-equilibrium initial data by prescribing a suitable profile for the scalar field. Therefore, in addition to the metric, we also evolve a massless scalar field, with equation of motion
\begin{equation}
\Box \phi = 0\,,
\end{equation}
 and stress tensor
 \begin{equation}
 T_{\alpha\beta}=\partial_\alpha \phi \partial_\beta \phi - g_{\alpha\beta} \frac{1}{2} g^{\gamma\delta} \partial_{\gamma} \phi \partial_{\delta} \phi\,.
 \end{equation}	
	
To carry out the evolution, we write the general spacetime metric $g_{\mu\nu}=\hat g_{\mu\nu}+\bar g_{\mu\nu}$, where $\hat g_{\mu\nu}$ is the metric of the Poincar\'e patch of AdS$_4$ (with the AdS length $L=1$) written as
\begin{equation}
\hat g = \frac{\rho^4}{(1-\rho^2)^2} \left( -dt^2 + \tfrac{4}{\rho^6}\,d\rho^2 + dx^2 + dy^2  \right),
\label{eq:AdS4comp}
\end{equation}
where $z=(1-\rho^2)/\rho^2$ is the usual AdS radial coordinate, and $x$, $y$ are the spatial boundary directions, which we take to be Cartesian directions with infinite range. In practice we compactify these directions as $x=\tan(\frac{\pi}{2}\bar x_1)$, $y =\tan(\frac{\pi}{2}\bar x_2)$.
 $\bar g_{\mu\nu}$ is a general deviation away from AdS$_4$, not necessarily small, and it satisfies Dirichlet boundary conditions at the AdS boundary. Similarly, we write the scalar field as
\begin{equation}
\phi = (1-\rho^2)^2\bar\phi\,,
\end{equation}
and the evolved variable $\bar\phi$ vanishes at the boundary of AdS$_4$. 

Following \cite{Bantilan:2020pay,Bantilan:2020xas}, we write the source functions  as
\begin{equation}
H_\mu = \hat H_\mu + (1-\rho^2)\bar H_\mu\,,
\label{eq:sourceH}
\end{equation}
where $\hat H_\mu$ are the source functions for AdS$_4$ in the coordinates of equation \eqref{eq:AdS4comp} and $\bar H_\mu$ are the actual evolved source functions; the power of $(1-\rho^2)$ in \eqref{eq:sourceH} has been chosen so that $\bar H_\mu\big|_{\rho=1}=0$ at the boundary of AdS.  Following the prescription of \cite{Bantilan:2020xas}, we fix our gauge such that near the boundary of AdS, we have
\begin{equation}
\begin{aligned}
\bar{H}^{(1)}_{t}&=\frac{3}{2} \bar{g}^{(1)}_{\text{$t$$\rho $}}\,, \nonumber\\
\bar{H}^{(1)}_{\rho}&=\frac{3}{2} \bar{g}^{(1)}_{ \rho \rho }\,,\nonumber\\
\bar{H}^{(1)}_{x}&=\frac{3}{2} \bar{g}^{(1)}_{ \rho x }\,,\nonumber\\
\bar{H}^{(1)}_{y}&=\frac{3}{2} \bar{g}^{(1)}_{ \rho y }\,,
\end{aligned}
\end{equation}
where
\begin{equation}
\begin{aligned}
\bar{H}_\mu&=(1-\rho^2)\bar{H}^{(1)}_\mu+O((1-\rho^2)^2)\,,\\ 
\bar{g}_{\mu\nu}&=(1-\rho^2)\bar{g}^{(1)}_{\mu\nu}+O((1-\rho^2)^2)\,,
\end{aligned}
\end{equation}
 near the boundary. The source functions are set to zero in the interior of the spacetime. We choose the same smooth transition functions as \cite{Bantilan:2020xas}  to fix the source functions everywhere in the spacetime. 
 
We construct time symmetric initial data by prescribing a Gaussian profile for the initial scalar field, 
\begin{equation}
\phi = (1-\rho^2)^3\,A\,e^{-\tilde r(\rho,\bar x_1,\bar x_2)^2/\Delta^2} ,
\end{equation}
where $A$ and $\Delta$ control the amplitude and the width of the Gaussian respectively, and
\begin{equation}
\tilde r(\rho,x,y)^2 = \tfrac{(\rho-\rho_0)^2}{a_\rho^2} + \tfrac{\bar x_1}{\alpha_1^2}+\tfrac{\bar x_2}{\alpha_2^2}\,.
\end{equation}
Here $\rho_0$ allows us to localize the initial Gaussian in the AdS radial direction, while the constant parameters $\{a_\rho,\alpha_1,\alpha_2\}$ control the shape along the boundary directions. Having specified an initial profile for the scalar field, the Hamiltonian constraint is solved as in \cite{Bantilan:2020xas}. We choose ``strong" data so that there is a trapped surface, with planar topology, in the initial data slice. 

We extract the stress energy tensor of the boundary CFT as in \cite{Bantilan:2020xas}. We find, 
\begin{align}
\langle T_{tt} \rangle &= \frac{1}{32\,\pi}\left( 6\,\bar g^{(1)}_{xx} + 6\,g^{(2)}_{yy}+\bar g^{(1)}_{\rho\rho}\right)\,,\\
\langle T_{tx}\rangle & = \frac{3}{16\,\pi}\,\bar g^{(1)}_{tx}\,,\\
\langle T_{ty}\rangle & = \frac{3}{16\,\pi}\,\bar g^{(1)}_{ty}\,,\\
\langle T_{xx} \rangle &= -\frac{1}{32\,\pi}\left(-6\,\bar g^{(1)}_{tt} + 6\,g^{(2)}_{yy}+\bar g^{(1)}_{\rho\rho}\right)\,,\\
\langle T_{xy} \rangle & = \frac{3}{16\,\pi}\,\bar g^{(1)}_{xy}\,,\\
\langle T_{yy} \rangle &= -\frac{1}{32\,\pi}\left(-6\,\bar g^{(1)}_{tt} + 6\,g^{(2)}_{xx}+\bar g^{(1)}_{\rho\rho}\right)\,.
\end{align}
 
Fig. \ref{fig:trace_conv_tests} shows a convergence test for the trace of the boundary stress tensor of the evolution presented in the main text, which shows first order convergence with the number of points along the holographic direction $N_{\rho}$.
\begin{figure}[h]
	\centering	
	\includegraphics[width=0.64\textwidth]{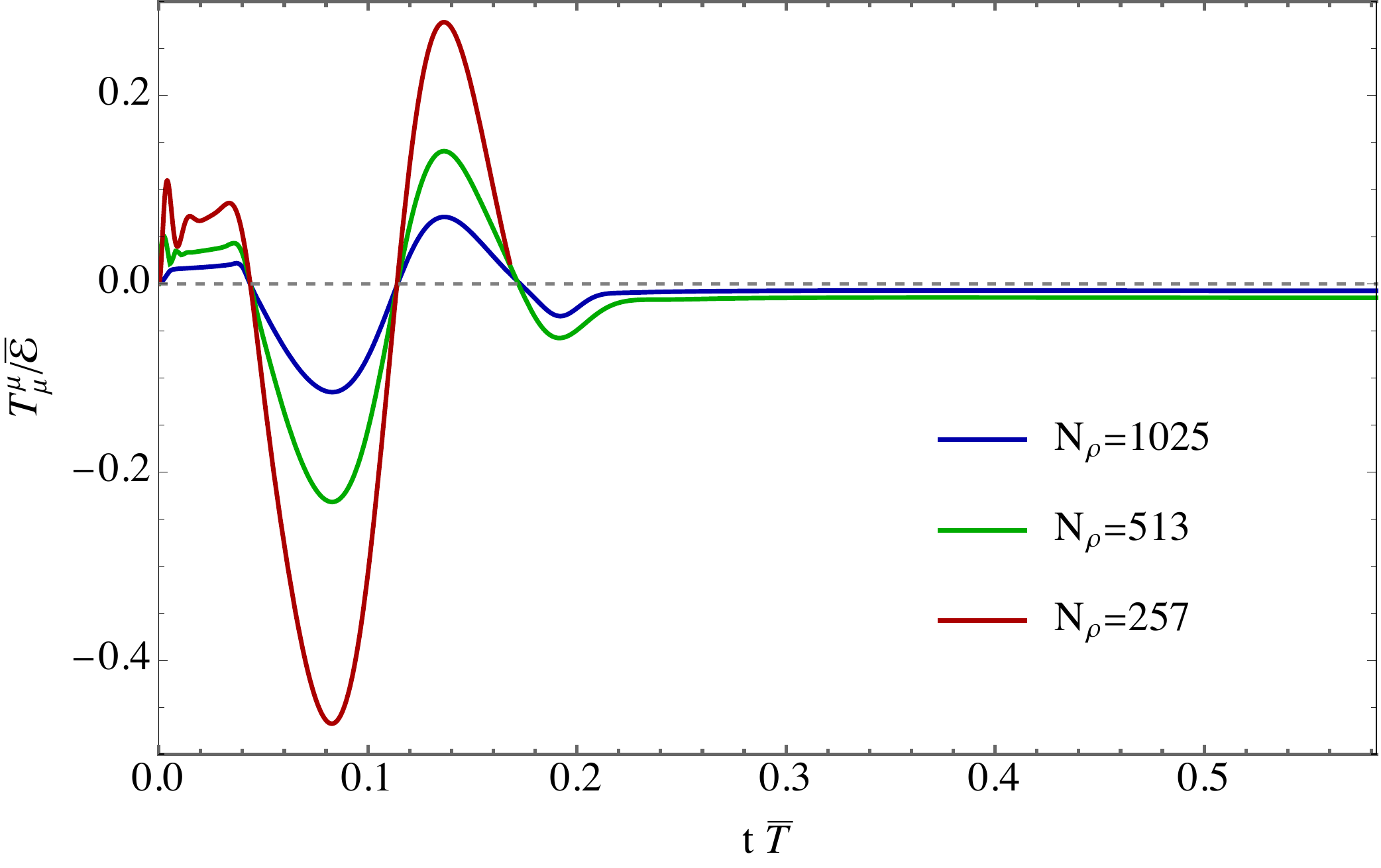}	
	\includegraphics[width=0.64\textwidth]{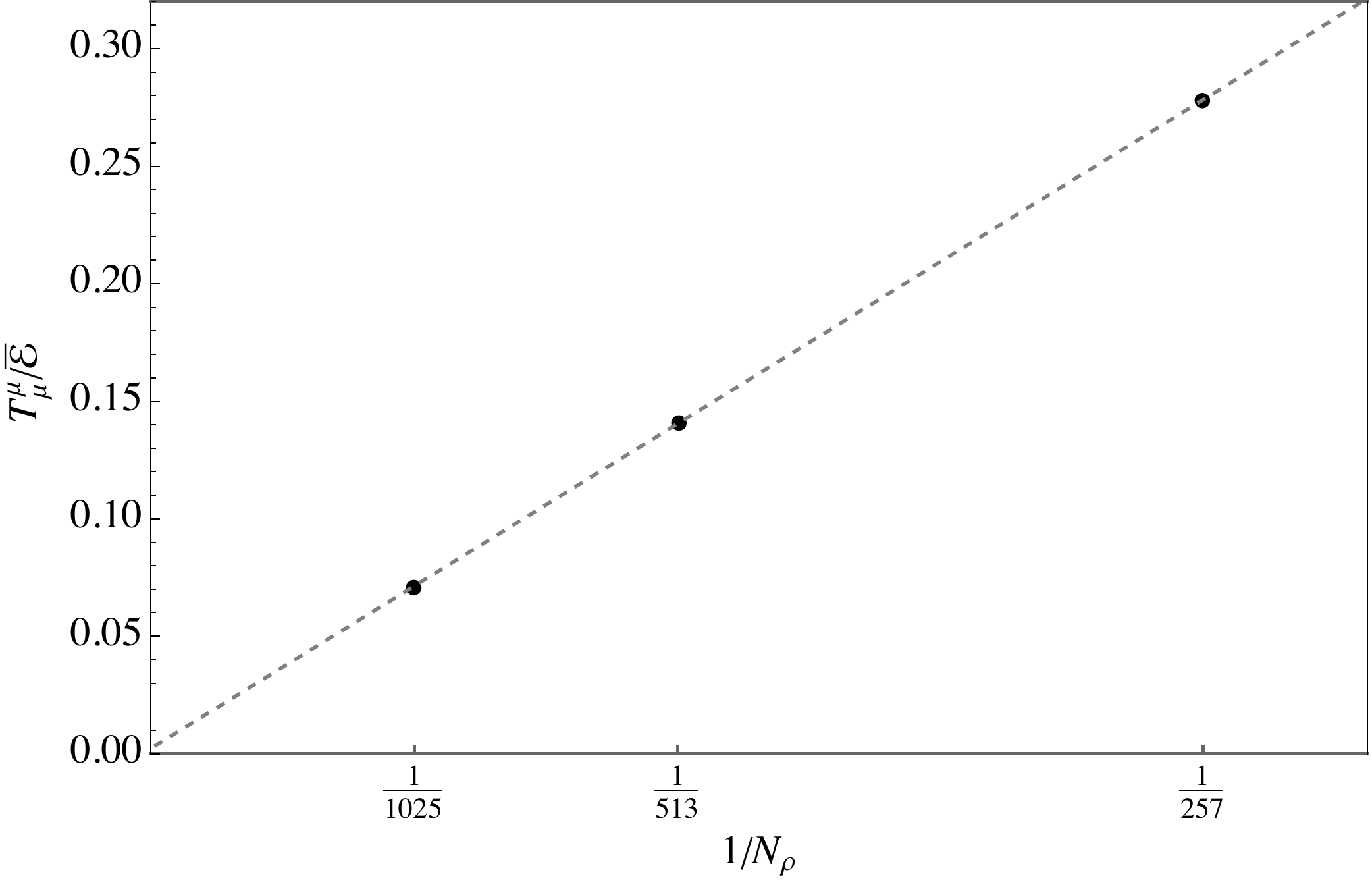}
	\caption{Convergence test for the trace. (Top) Trace at the center of the domain $\{x,y\}\overline{T}\simeq \{0,0\}$ as a function of time, for three different resolutions along the holographic direction $N_{\rho}=257,513,1025$. The data for the $N_{\rho}=257$ case is only available up to $t\overline{T}\simeq 0.17$. (Bottom) Trace as a function of $1/N_{\rho}$ at a fixed time $t\overline{T}\simeq 0.14$, at the center of the domain. The dashed line is a straight line fit that shows first order convergence. }
	\label{fig:trace_conv_tests}
\end{figure}

A convergence test of the conservation equation of the stress tensor for our code was presented in \cite{Bantilan:2012vu}, and we do not perform this convergence test in our case. 
\begin{figure}[h]
	\centering	
	\includegraphics[width=0.64\textwidth]{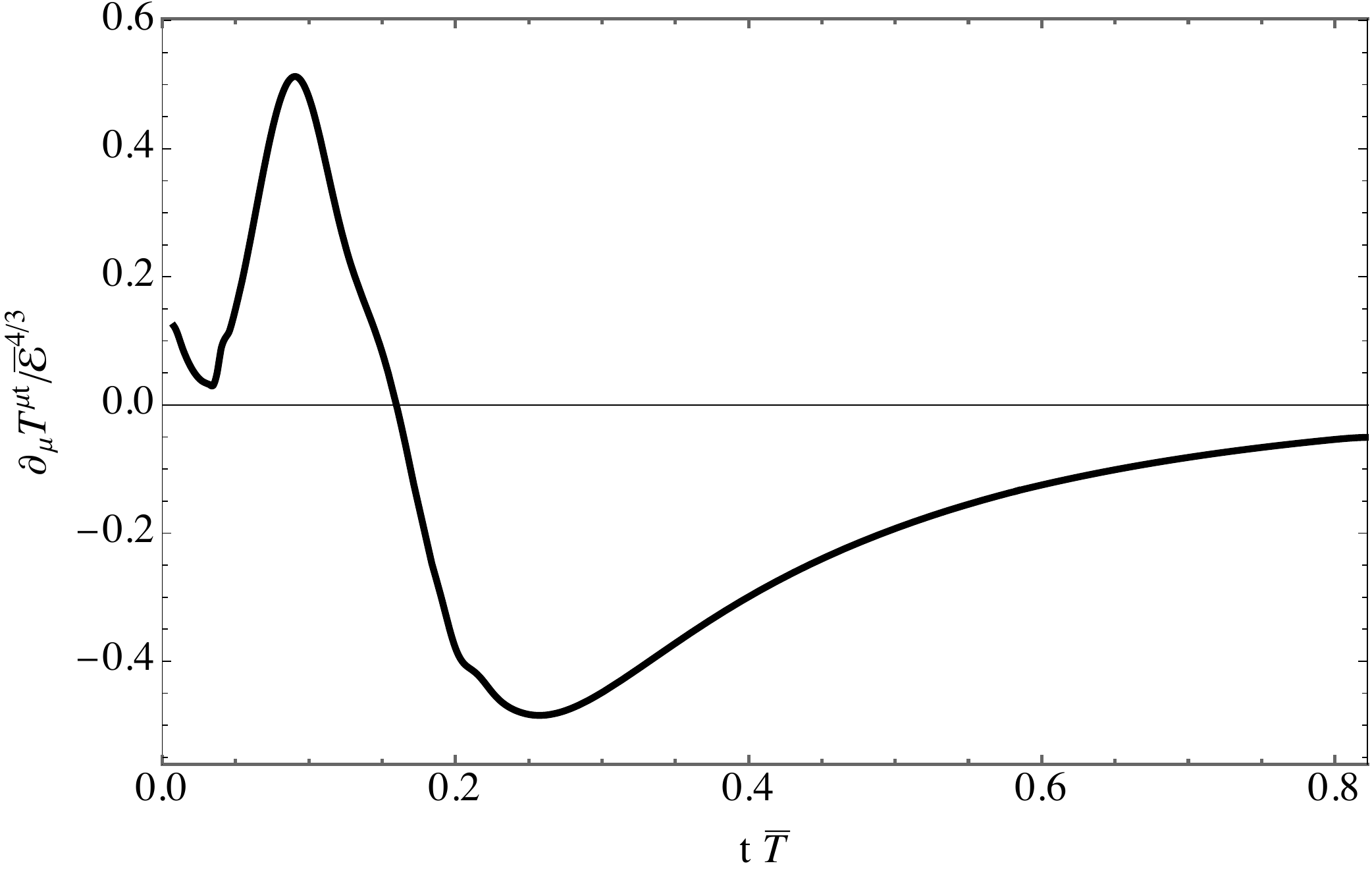}	
	\caption{Temporal component of the stress tensor conservation equation, $\partial_{\mu} T^{\mu t}$, at the center of the domain as a function of time.}
	\label{fig:conservation_stress_tensor}
\end{figure}
Fig. \ref{fig:conservation_stress_tensor} shows the temporal component of the conservation equation for the stress tensor, $\partial_{\mu} T^{\mu t}$, for our highest resolution case $N_{\rho}\times N_x\times N_y=1025\times129\times129$. We present the result at the center of the domain because the numerical error is maximal there. For the other components, $\partial_{\mu} T^{\mu x}$, $\partial_{\mu} T^{\mu y}$, the error is smaller. 

We use the numerical error of the conservation of the stress tensor, Fig. \ref{fig:conservation_stress_tensor}, to provide an estimate of the numerical error in the first order terms of the hydrodynamic expansion. Schematically, the first order terms are given by the shear viscosity times derivatives of the velocity, so we estimate the error as
\begin{equation}
\frac{T^{1st}_{xx}}{T^{id}_{xx}} \sim \frac{\eta \, \partial_x v_x}{\overline{\mathcal E}/2} \sim \frac{\eta \, \partial_{\mu} T^{\mu t}/(2\overline{\mathcal E})}{\overline{\mathcal E}/2} \,.  
\label{errorestimation}
\end{equation}
 We estimate $T^{id}_{xx} \sim \overline{\mathcal E}/2$ and $\partial_x v_x \sim \partial_{\mu} T^{\mu t}/(2\overline{\mathcal E})$. In the last expression, by dividing by $\overline{\mathcal E}$ we obtain the correct units, and the factor of two is because at the center the $\partial_{x} T^{x t}$ and $\partial_{y} T^{y t}$ terms contribute equally and from the actual data we check that the $\partial_{t} T^{t t}$ term is much smaller. 
For times around the hydrodynamization time $t\overline{T}\simeq 0.5$, the right hand side of (\ref{errorestimation}) gives an error of the order
\begin{equation}
\frac{T^{1st}_{xx}}{T^{id}_{xx}}  \sim 2\%\, .
\label{errorestimation2}
\end{equation}
A numerical error of the order of $\sim 2 \%$ is of the same order of the first and second order terms around times $t\overline{T}\simeq 0.5$, see Fig. \ref{Constitutive_relations_Frame2}. So, all the statements in this paper about the applicability of hydrodynamics for quantities beyond this time must be considered only up to numerical errors. 

The change from the Landau frame to the causal frame is specially affected by the error in the conservation equation of the stress tensor. This is because the first order terms in the change of frame expression are precisely the same terms as in the ideal conservation equation of the stress tensor. For this reason, we subtract by hand the error of the conservation of the stress tensor in the change of frame to obtain the initial data for the BDNK evolutions.  

%%%%%%%%%%%%%%%%%%%%%%%%%%%%%%%%%%%%%%%%%%%%%%%%%%%%%%%%%
\section{Details on the dynamical evolutions}
\label{CCappedix}
%%%%%%%%%%%%%%%%%%%%%%%%%%%%%%%%%%%%%%%%%%%%%%%%%%%%%%%%%

In this appendix we provide more details about the dynamical evolutions presented in the main text.

Fig. \ref{Constitutive_relations_Frame2} shows the ratios $T_{xx}^{n}/T_{xx}^{ideal}$, $n=1,2$, of the constitutive relations (\ref{constitutive0sheartensor0}) evaluated at $\{x,y\}\overline{T}\simeq \{0.12,0\}$ as a function of time for the holographic solution. This shows explicitly the validity of the hydrodynamic expansion. Alternatively, the constitutive relations can also be compared with the microscopic stress tensor values, to assess if the constitutive relations describe the microscopic evolution. This is shown in Fig. \ref{fig:constitutive_relations_holography}. 
On general grounds, we would expect that if the system is in the hydrodynamic regime, then the gradient corrections should improve the ideal description. 
However, in our system, at the hydrodynamization time $t\overline{T}\simeq 0.5$ first and second order terms do not improve the ideal description of the microscopic solution. 
As stated in the main text, this might be explained by our choice initial data.
We expect that at later times viscous descriptions
would improve the ideal result. However, at later times the numerical error (discussed in appendix B) does not allow us to make conclusive statements in this regard. Also, possibly for the same reason, ideal terms provide a good description of the system even before the hydrodynamization time $t\overline{T}\simeq 0.5$. 
The microscopic $T_{xx}$ presented in Fig. \ref{fig:constitutive_relations_holography} presented a numerical oscillation of a cero mode (homogeneous). This numerical oscillation converges to zero similarly to Fig. \ref{fig:trace_conv_tests}, and we removed it by hand in Fig. \ref{fig:constitutive_relations_holography}.

\begin{figure}[t]
	\centering	
	\includegraphics[width=0.64\textwidth]{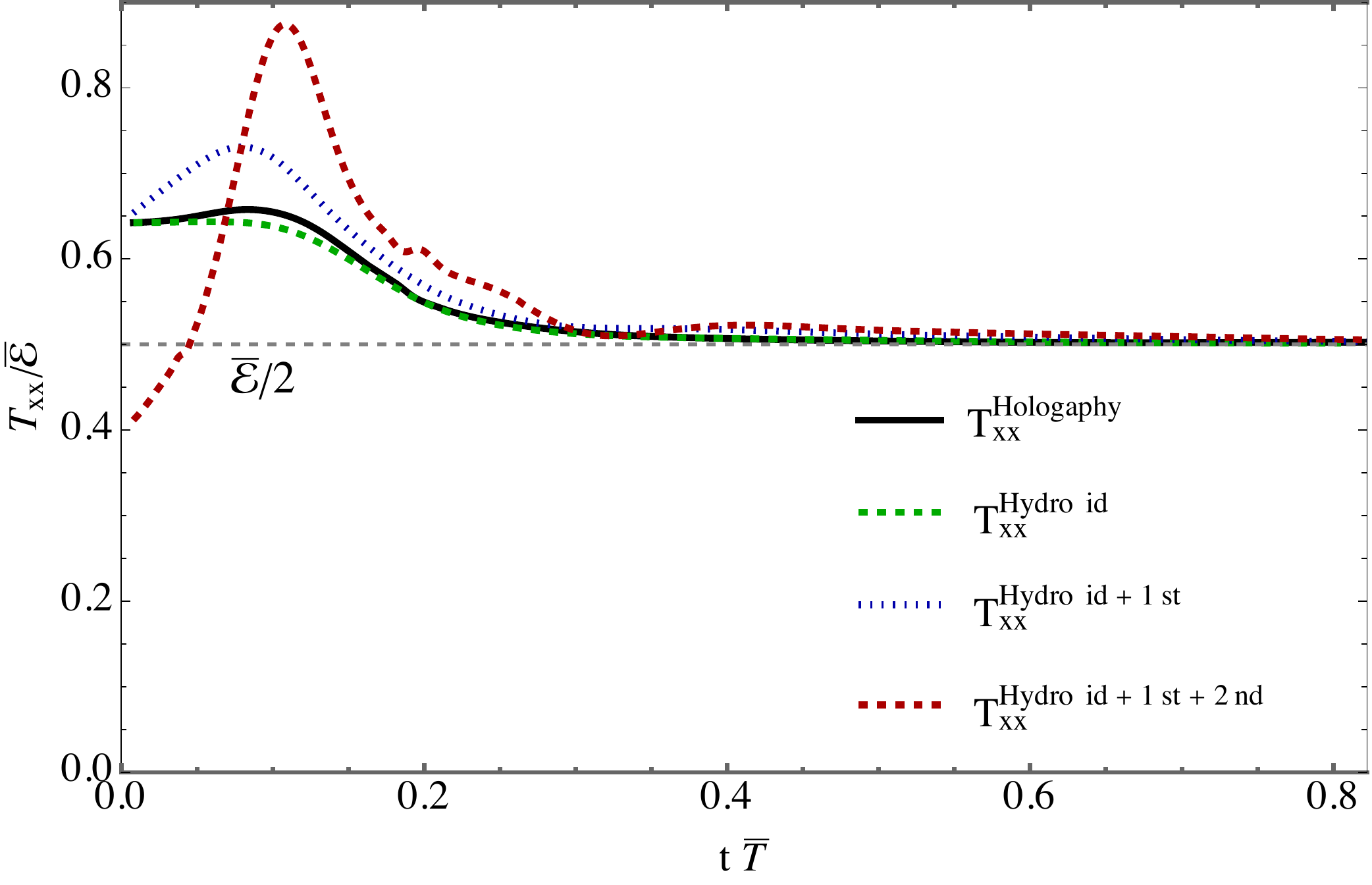}	
	\caption{$T_{xx}$ component of the holographic stress tensor at $\{x,y\}\overline{T}\simeq\{0.12,0\} $ as a function of time, solid black line. We include the values of $T_{xx}$ obtained by evaluating the constitutive relations (\ref{constitutive0sheartensor0}) up to ideal terms (dashed green), first order terms (dotted blue) and second order terms (dashed red).}
	\label{fig:constitutive_relations_holography}
\end{figure}

\begin{figure}[h]
	\centering
    \includegraphics[width=0.64\textwidth]{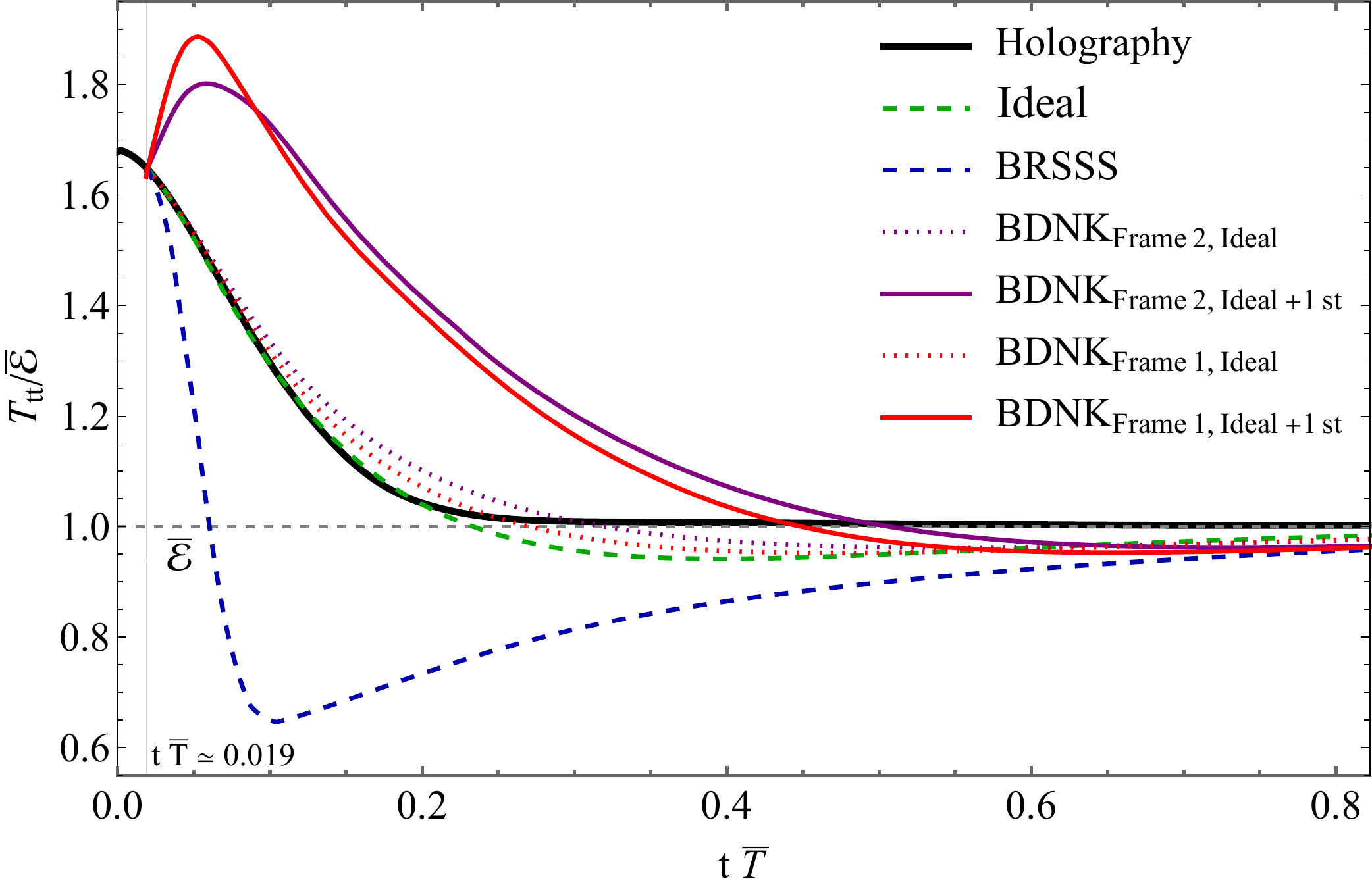}	
   	\includegraphics[width=0.64\textwidth]{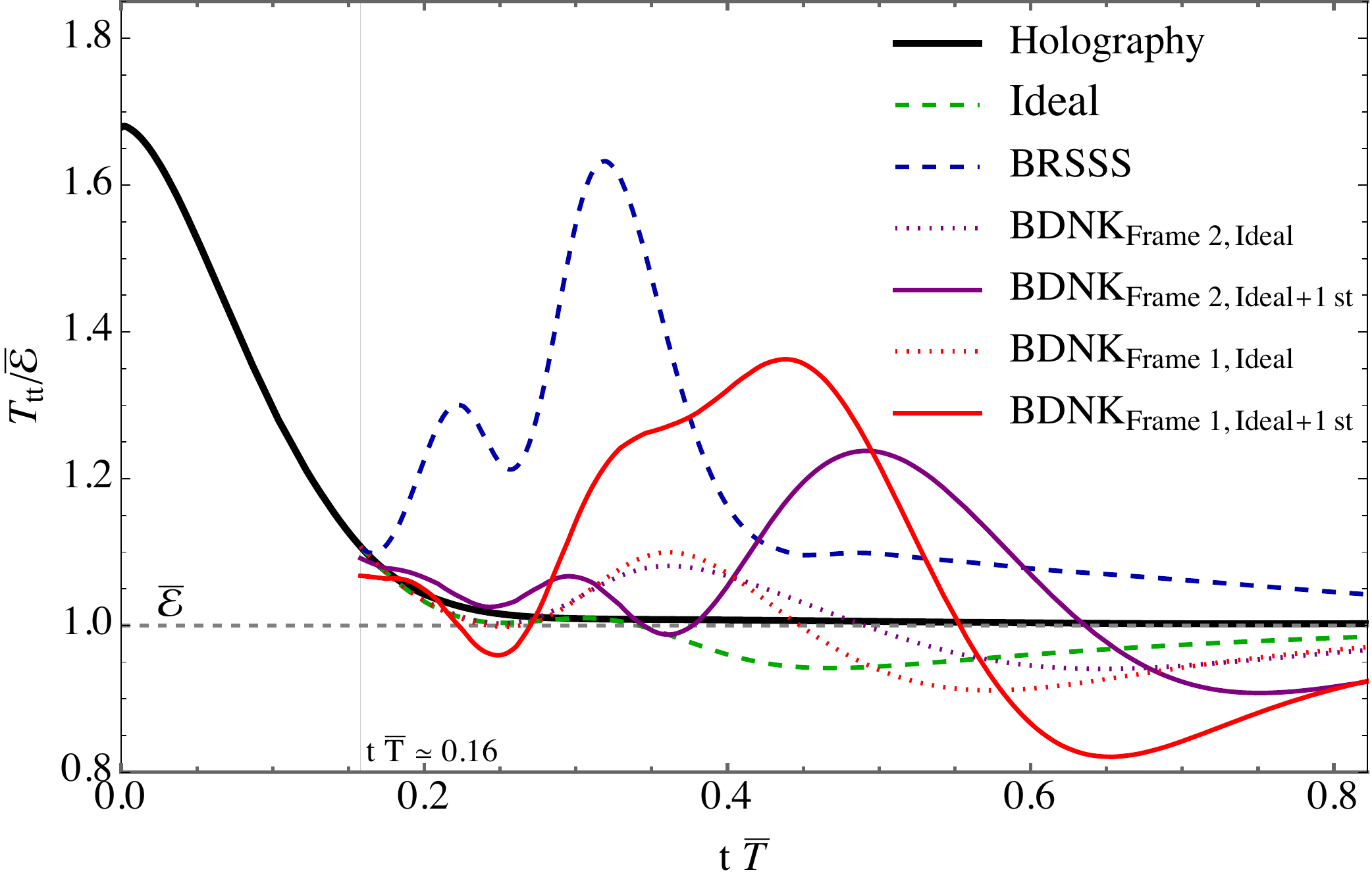}	
	\includegraphics[width=0.64\textwidth]{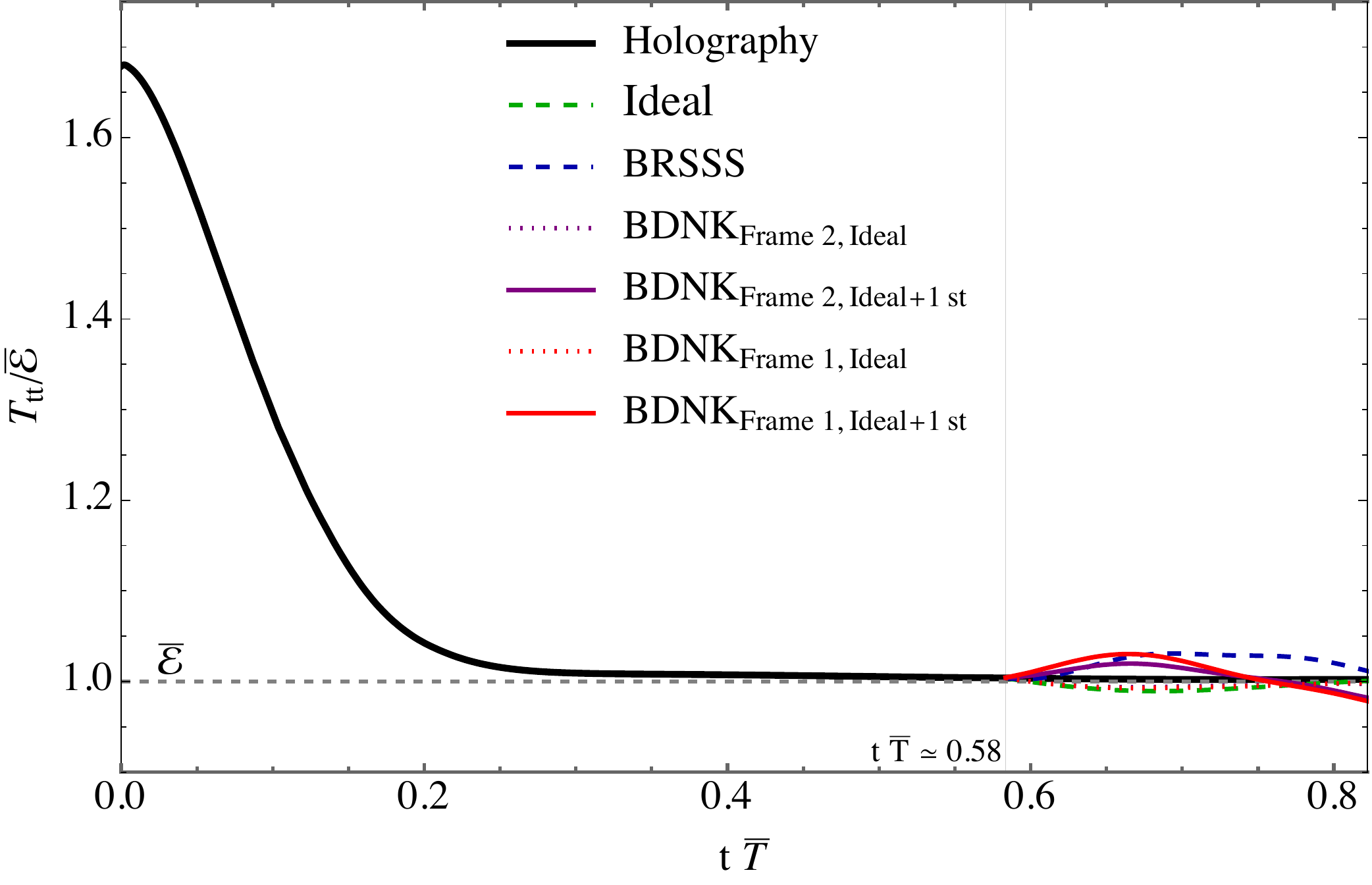}	
	\caption{Energy density in the lab frame, $T_{tt}$, at the center of the domain $\{x,y\}\overline{T}=\{0,0\}$ as a function of time. Black continuous line corresponds to the microscopic holographic solution. We include the results of the hydrodynamic evolutions initialised with holographic data at $t\overline{T}\simeq 0.019, 0.16, 0.58$, from top to bottom. For the BDNK evolutions, we include the result of using only the 0th order terms (the ideal part) in (\ref{eq:tmunu11}), in dotted lines. We also include the evolutions of BDNK in the frame 1.}
	\label{Evolutions_center_ideal}
\end{figure}
Fig. \ref{Evolutions_center_ideal} shows the same evolutions as in Fig. \ref{fig:off_center_evolutions} but at an off center location $\{x,y\}\overline{T}=\{0.17,0\} $. The conclusions are similar to those obtained from Fig. \ref{fig:off_center_evolutions}. If the gradients are large (top and middle), the hydrodynamic evolutions provide a description that differs from the microscopic one, and are also different among them, as all theories have different UV behaviors. At late times (bottom), when gradients are small, the hydrodynamic theories provide a better description of the system, matching the microscopic theory within a maximum deviation in the domain of a $\sim 2\%$ in the energy density.

Fig. \ref{Evolutions_center_ideal} also includes the evolutions of the BDNK equations in frame 1, and we discuss now the differences of evolving the BDNK equations in the different causal frames \eqref{frames}. First, recall that $a_1$ and $a_2$ are multiplying first order terms in \eqref{eq:tmunu11}. This means that the larger these values, the larger are those first order terms. So, for frame 1 these terms are larger than for frame 2. This is the reason why in the main text we focus on the frame 2, for which the values for $a_1$ and $a_2$ are closer to the smallest ones allowed by causality. 
The size of the constants $a_1$ and $a_2$ will be relevant when changing from the Landau frame to the causal frame, when evolving the equations, and also when going back to the Landau frame. We observe that the size of these first order terms (the ones proportional to $a_1$ and $a_2$) in the microscopic solution is typically small at all times, even far from equilibrium. On the other hand, in the BDNK simulations in the far from equilibrium regime, these terms become large, even of order one, see Fig. \ref{Evolutions_center_ideal}  (top and middle). 
Moreover, the BDNK evolutions in the different frames in the far from equilibrium region may provide considerably different results, see Fig. \ref{Evolutions_center_ideal}  (middle). 
Finally, in Fig. \ref{Evolutions_center_ideal} (bottom) we see that the difference between frame 1 and 2 is compatible with second order terms, in accordance with the fact that the system has hydrodynamized.

\end{appendix}

\bibliographystyle{JHEP}
\bibliography{refs_BDNK}

\end{document}